\documentclass{elsarticle}
\usepackage{color,graphicx,subfigure}
\usepackage{listings}
\usepackage{algorithm}
\usepackage{algpseudocode}
\usepackage{amsmath}

\usepackage{hyperref}

\makeatletter
\def\ps@pprintTitle{%
 \let\@oddhead\@empty
 \let\@evenhead\@empty
 \def\@oddfoot{}%
 \let\@evenfoot\@oddfoot}
\makeatother

\usepackage{tikz}
\usetikzlibrary{decorations.pathreplacing,calc}
\newcommand{\tikzmark}[1]{\tikz[overlay,remember picture] \node (#1) {};}

\newcommand*{\AddBraceNote}[4]{%
    \begin{tikzpicture}[overlay, remember picture]
        \draw [decoration={brace,amplitude=0.5em},decorate,ultra thick,black]
            ($(#3)!(#1.north)!($(#3)-(0,1)$)$) --
            ($(#3)!(#2.south)!($(#3)-(0,1)$)$)
                node [align=left, text width=2.5cm, pos=0.5, anchor=west] {$\quad$#4};%
    \end{tikzpicture}%
}%
\newcommand*{\AddArrowNote}[4]{
    \begin{tikzpicture}[overlay, remember picture]
        \draw [->,decorate,ultra thick,black]
            ($(#2)!(#1)!($(#2)-(0,1)$)$) --
            ($(#3)!(#1)!($(#3)-(0,1)$)$);
        \node [align=left,  text width=2.5cm, anchor=west] (a) at ($(#2)!(#1)!($(#2)-(0,1)$)$)   {$\quad$#4};
    \end{tikzpicture}
}%

\begin{document}

\begin{frontmatter}

\title{Locality-aware parallel block-sparse matrix-matrix multiplication using the
  Chunks and Tasks programming model}

\author{Emanuel H. Rubensson}
\ead{emanuel.rubensson@it.uu.se}
\author{Elias Rudberg}
\ead{elias.rudberg@it.uu.se}
\address{Division of Scientific Computing, Department of Information Technology, Uppsala University, Box 337, SE-751 05 Uppsala, Sweden}

\begin{abstract} 
  We present a method for parallel block-sparse matrix-matrix
  multiplication on distributed memory clusters. By using a quadtree
  matrix representation, data locality is exploited without prior
  information about the matrix sparsity pattern.  A distributed
  quadtree matrix representation is straightforward to implement due
  to our recent development of the Chunks and Tasks programming model
  [Parallel Comput. 40, 328 (2014)].  The quadtree representation
  combined with the Chunks and Tasks model leads to favorable weak and
  strong scaling of the communication cost with the number of
  processes, as shown both theoretically and in numerical experiments.

  Matrices are represented by sparse quadtrees of chunk objects.  The
  leaves in the hierarchy are block-sparse submatrices. Sparsity
  is dynamically detected by the matrix library and
  may
  occur at any level in the hierarchy and/or within the submatrix
  leaves. In case graphics processing units (GPUs) are available, both
  CPUs and GPUs are used for leaf-level multiplication work, thus
  making use of the full computing capacity of each node.

  The performance
  is evaluated for matrices with
  different sparsity structures, including examples from electronic structure calculations.
  Compared to methods that do not exploit data locality, our
  locality-aware approach reduces communication significantly,
  achieving essentially constant communication per node in weak scaling tests.

\end{abstract}

\begin{keyword}
parallel computing \sep
sparse matrix-matrix multiplication \sep
scalable algorithms \sep
large-scale computing \sep
graphics processing units
\end{keyword}

\end{frontmatter}

\section{Introduction}
Sparse matrix-matrix multiplication, sometimes referred to as SpGEMM,
is a key operation in large-scale electronic structure calculations
based on for example Hartree--Fock or Kohn--Sham density functional
theory~\cite{DBowler12}. Sparse matrix-matrix multiplication is used
in particular in polynomial expansion~\cite{pur-pm} and minimization
methods~\cite{dmm-lnv} to compute the density matrix.  Such methods
are used in a number of electronic structure codes such as {\sc
  Conquest}~\cite{conquestGillan200714}, {\sc
  CP2K}~\cite{cp2k-linearscaling}, {\sc Ergo}~\cite{linmemDFT}, {\sc
  FreeON}~\cite{FreeON}, {\sc Honpas}~\cite{honpas}, {\sc
  Onetep}~\cite{onetep}, and {\sc LATTE}~\cite{LATTE-jcp-2012} to
achieve a computational cost that increases only linearly with system
size.  The matrix sparsity varies from tens
to thousands of nonzeros per row depending on the underlying model and
the basis set used. It is often beneficial to use a block-sparse data
structure. The optimal block size depends on the model and on the
order of the matrix rows and columns.  The present work is mainly
motivated by Hartree--Fock and Kohn--Sham density functional theory
calculations using Gaussian basis sets in which the matrices have up
to thousands of nonzero elements per row and a priori unknown sparsity
patterns~\cite{sparsity-JCC:JCC21723, linmemDFT}.  This work is also
relevant for the general parallel SpGEMM problem as no application
specific information such as atomic positions is built into the
presented method.

\begin{figure}
  \begin{center}
    \includegraphics[width=0.3\textwidth]{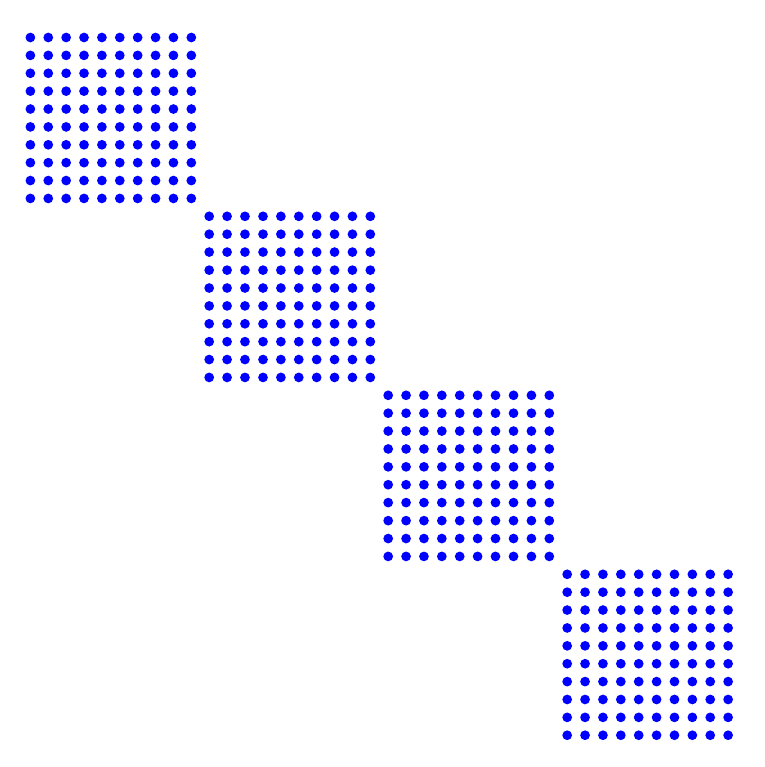}
    \includegraphics[width=0.3\textwidth]{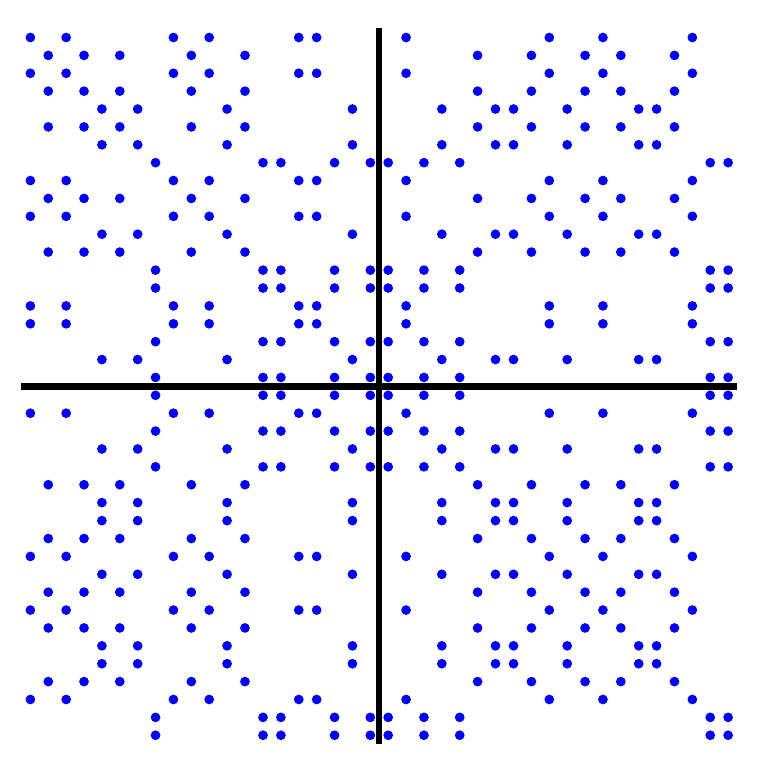}
  \end{center} 
  \caption{A trivial example of possible effects of random permutation
    to load balance static work and data distribution. Left: Example
    matrix for which we want to compute the square on 4 identical
    compute nodes. The best distribution of data and work is obvious
    and leads to perfect load balance and no communication of matrix
    elements. Right: Random permutation of the matrix columns and rows
    and a two-dimensional data decomposition indicated by solid black
    lines.  Although the workload may still be roughly load balanced,
    communication of matrix elements is now
    needed.  \label{fig:random_destruction}}
\end{figure}

Algorithms based on dense matrix-matrix multiplication are generally
considered attractive because of the existence of efficient linear
algebra libraries, e.g.~\cite{Goto-matrix-mul, whaley04}, and
parallelization through e.g.~Cannon's algorithm or
SUMMA~\cite{summa}. Parallel sparse and block-sparse matrix-matrix
multiplication has received less attention and represents a greater
challenge, particularly when the nonzero pattern is not known in
advance.  Nevertheless, several parallel sparse matrix-matrix
multiplication methods have been presented.  Several methods assume
some a priori knowledge about the input matrix sparsity structure and
use that knowledge to improve
performance~\cite{conquest-sparsematrix,Challacombe-sparsematrix,
  onetep-sparsematrix,WeberEtAlMidpointMmul2015}.  Other methods
require beforehand knowledge of the computational pattern or the
sparsity structure of the output matrix, requiring a preparatory
symbolic multiplication step before the actual parallel computation
can start~\cite{Akbudak2014, Ballard2016}.  Here, we will focus on the
general case where no a priori knowledge about the structure is
assumed, and no symbolic multiplication step is needed. Recent methods
for this general case are first employing a random permutation of the
rows and columns of the matrix to destroy any structure in the
sparsity pattern and \emph{decrease} data
locality~\cite{BulucGilbert2012, Borstnik2014}.
The goal of this
maneuver is to obtain about the same density of nonzero elements
everywhere in the matrix. Then, a static distribution of work and data
is used in the same way as for dense matrices, but with the local
block-block multiplies replaced by sparse products. This random permutation approach prevents load imbalance, but the obvious
drawback is that the possibility to exploit the nonzero structure to
reduce communication or make efficient use of the memory hierarchy is
spoiled, see Figure~\ref{fig:random_destruction} for a trivial yet
illustrative example.  On the other hand, such exploitation is
difficult to achieve since it requires that the mapping of data and
work to physical resources is performed dynamically during the
calculation~ \cite{communication_optimal_2}.
We believe that the difficulties are mainly associated with the
programming model used to tackle the problem.  While conventional
programming models like message passing protocols work well for static
distribution of work and data, they are inconvenient if you want to
distribute data and work dynamically.  The programmer has to make
decisions about where data should be located, where every piece of
work should be executed, and see to it that data is communicated as
needed.

Recently, we proposed a new programming model named Chunks and Tasks,
designed to work well for algorithms with dynamic work and
data~\cite{chunks-and-tasks}.
We describe and analyze in this article an approach to parallel
block-sparse matrix-matrix multiplication based on a hierarchical
quaternary tree (quadtree) representation implemented using the Chunks
and Tasks programming model. The method is locality-aware in the sense
that it is able to exploit a priori unknown structure in the sparsity
pattern to reduce communication and thereby improve performance. The
library used for performance evaluation here is a further developed
version of the code that was briefly described and used for test
calculations in \cite{chunks-and-tasks}.

This article is organized as follows: in Section~\ref{sec:model}, we
briefly discuss the Chunks and Tasks programming model. Our new matrix
library based on Chunks and Tasks is presented in
Section~\ref{sec:chtml}.  In the present work, the Chunks and Tasks
matrix library is used together with a block-sparse leaf matrix type,
described in Section~\ref{sec:leaf_matrix_types}.
An analysis of the computational costs due to the quadtree
representation is given in Section~\ref{sec:quadtree-effects},
followed by results of test calculations in
Section~\ref{sec:performance} and concluding remarks in
Section~\ref{sec:conclusions}.

\section{Programming model}\label{sec:model}
Our block-sparse matrix library has been implemented using the Chunks
and Tasks programming model~\cite{chunks-and-tasks}. In Chunks and
Tasks the programmer writes her program in terms
of small pieces of data and work, chunks and tasks, respectively. The
programmer is responsible for dividing work and data into smaller
pieces but not for the mapping of work and data onto physical
resources. The programmer need not worry about message passing, all
communication is handled by the Chunks and Tasks library. The
programmer does neither have to worry about race conditions nor
non-deterministic behavior.
The computation is driven by the registration of tasks,
similarly to other task-based models. Recursive nesting of tasks is
allowed, i.e.~during task execution new tasks can be registered as in
for example Cilk~\cite{cilk}, Scioto~\cite{scioto},
SuperGlue~\cite{superglue}, and XKaapi~\cite{xkaapi}. This is
important for scalability of dynamic algorithms, since otherwise only
a single process can generate new tasks, or multiple processes
generate predetermined (static) task graphs.

A key feature of the Chunks and Tasks model is that abstractions are
not only provided for work but also for data.  The Chunks and Tasks
library takes care of the distribution of both work and data. The user
creates data objects called chunks. The transfer of responsibility of
such a chunk object to the runtime library is referred to as
registration of a chunk; in return, the user gets an identifier that
can be used to specify dependencies later on.  After the point of
registration the chunk object is read-only.  This is in a way similar
to e.g.~Linda~\cite{Carriero1994633} and Concurrent
Collections~\cite{CnC} that also have a ``space'' to which you can add
a piece of data and later retrieve it, possibly on another process. A
key difference is that in Linda and Concurrent Collections the
identifier is chosen by the application programmer whereas in Chunks
and Tasks, the identifier is chosen by the runtime library. On one
hand, being able to choose identifiers makes it possible for a process
or task to ask for data without any prior communication or interaction
whatsoever with the process or task that registered the data.
On the other hand, the fact that the runtime library does not control
the identifiers means that inconsistencies can be introduced
(supposedly unintentionally) where for example several different
pieces of data with the same identifier exist, e.g.~on distant nodes
in a cluster.
Perhaps of even greater importance is that such a model makes it
difficult for the runtime library to make data available efficiently.
Any process may ask for any piece of data at any time possibly
without any information being available locally about the location of
the piece being asked for.  This stands in contrast to Chunks and
Tasks where the library for example can store information about the
location in the chunk identifier.  In this way, Chunks and Tasks,
by imposing appropriate restrictions, makes life easier both for the
application programmer and the runtime library developer.

\subsection{Library implementations}\label{subsec:cht-lib-impl}

A Chunks and Tasks program can be compiled, linked and executed with
any Chunks and Tasks runtime library implementation. We will in our
performance evaluation use the publicly available Chunks and Tasks
library CHT-MPI~\cite{cht-mpi, chunks-and-tasks}, which is written in
C++ and uses the Message Passing Interface (MPI) for communication
between computational nodes, enabling Chunks and Tasks programs to run
on distributed-memory clusters.

The CHT-MPI implementation uses work stealing to distribute tasks, see
e.g.~\cite{BlumofeAndLeiserson1999}, meaning that there is no central
"master" node responsible for all scheduling. Each worker process is
responsible for its own set of tasks, and steals work from some other
(randomly selected) worker when it has no work left. For recursive
algorithms operating on hierarchical data structures, some tasks
recursively generate new tasks leading to a tree of tasks, and each
worker process effectively executes its own local part of that tree.
Work stealing always occurs as high up as possible in the local task
tree of the victim process.

Besides distributing tasks, the runtime library must also make sure
that the necessary input data is available for each task. Thanks to
the possibility of storing the MPI rank of the owner process in the
chunk identifier, this becomes straightforward: when a particular
chunk is needed as input to a task, the library implementation simply
inspects the chunk identifier to find out from which worker the data
should be fetched. In this way data can be made available efficiently,
without need for any central authority storing information about the
location of all chunks.

Each chunk object is by default owned by the worker process that
created that chunk. This has the advantage that no communication is
needed to create a chunk, and temporary chunks used within a local
part of the task tree can often be reused directly without need for
any communication, since each worker processes its own local part of
the task tree. CHT-MPI also implements a chunk cache for each worker
process, meaning that if the same chunk is needed multiple times it
only needs to be fetched the first time. Thus, for Chunks and Tasks
programs corresponding to recursive algorithms operating on
hierarchies of chunks, data re-use happens automatically. Note that
the data distribution is determined dynamically and follows from the
work stealing distribution of tasks. The distribution of chunks among
worker processes will therefore in general be different for different
runs of the same program.

For the reasons outlined above, CHT-MPI can be used to achieve
scalable parallelization for Chunks and Tasks programs; both work and
data is distributed dynamically without need for any "master" node
that all workers must communicate with. From an application
programmer's point of view, what is needed to make use of these
features is to express the program using hierarchical representations
and recursive algorithms. As will be seen in the following section, a
quadtree-based representation is a natural way to achieve this for
matrix operations. See~\cite{cht-mpi, chunks-and-tasks} for more
information about CHT-MPI.

\section{Quadtree representation of matrices in the Chunks and Tasks model}\label{sec:chtml}
Hierarchical data structures based on a two-dimensional block
decomposition of the matrix at each level in a hierarchy have both
been used to block for the memory hierarchy in dense matrix
computations \cite{ChatterjeeMatMul2002_A, FrensMatMul1997_B,
  recursive_dense2004} and to avoid operations on zero elements (or
entire submatrices that are zero) in sparse matrix computations
\cite{quadtreeWise1984}.
Quadtree representations have also been advocated for
simplicity and expressiveness in particular leading
to
ease of programming for multiprocessing (shared memory) environments
\cite{Dinh_1999, lugowski2014, m-rrs, quadtreeWise1984,
  ChatterjeeMatMul2002_A, FrensMatMul1997_B} and straightforward
exploitation of symmetry \cite{m-rrs}.
As will be shown in Section~\ref{sec:quadtree-effects}, the quadtree
representation is in principle also appropriate for distributed
representation of sparse matrices on computer clusters. The caveat is
that a distributed sparse quadtree representation is difficult to
implement in conventional programming models, especially if a priori
unknown sparsity patterns are to be handled efficiently.  In this
section, we describe how such a sparse matrix quadtree representation
can be straightforwardly implemented in the Chunks and Tasks
programming model~\cite{chunks-and-tasks}.

In our Chunks and Tasks matrix library, matrices are represented by
sparse quadtrees of chunks.  At the lowest level in the
hierarchy, different leaf matrix representations, for example dense or
sparse, may be used.  In this work we will focus on regular
matrix-matrix multiplication on the form $C = AB$ and the symmetric
matrix square operation $C=A^2$, where $A$ and therefore also $C$ is
symmetric and only the upper triangles of $A$ and $C$ are stored.
The sparse symmetric matrix square is a key operation and a major
computational challenge in linear scaling electronic structure
calculations.
In all task type implementations, sparsity is dynamically exploited at
all levels in the hierarchy by skipping operations on zero
submatrices, which are represented by NIL chunk identifiers. Regular
matrix-matrix multiplication without transpose $C = AB$ and matrix
addition $C = A+B$ are illustrated as pseudo-code in
Algorithms~\ref{alg:pseudo_multiply} and~\ref{alg:pseudo_add},
respectively. For all task types, at the lowest level in the
hierarchy, the corresponding functionality of the leaf matrix library
is used, while at higher levels a straightforward implementation for
the two by two case is used, see e.g. lines 7-14 in
Algorithm~\ref{alg:pseudo_multiply}. Checking for NIL chunk
identifiers corresponds to the if statement on line 2 in each of
Algorithms~\ref{alg:pseudo_multiply} and~\ref{alg:pseudo_add}. In
practice, the Chunks and Tasks C++ interface requires a regular
execute function used when all input chunks are available, and a
fallback execute used when some of the input chunk objects cannot be
constructed due to NIL chunk identifiers. This is convenient from a
programmer's point of view since the if statement on line 2 is checked
by the runtime library and the appropriate function, regular or
fallback execute, is called automatically. This also allows for
compile-time type checking and programming errors such as attempts to
access nonexisting chunks are not possible.
We list and describe below all chunk and task types that are needed
for the $C=AB$ and $C=A^2$ operations.

\begin{algorithm}[h!]
\begin{algorithmic}[1]
  \State \textbf{input:} $A, B$
  \If{$A$ not NIL \textbf{and} $B$ not NIL}
  \If{lowest level}  \tikzmark{top}
  \State $X$ = leafMatrixMultiply($A$, $B$)
  \State $C$ = registerChunk($X$)
  \Else
  \For{$m = 1, 2$}
  \For{$n = 1, 2$}
  \State $Y_1$ = registerTask(multiply, $A_{m1}$, $B_{1n}$)
  \State $Y_2$ = registerTask(multiply, $A_{m2}$, $B_{2n}$)
  \State $C_{mn}$ = registerTask(add, $Y_1$, $Y_2$)
  \EndFor
  \EndFor
  \State $C$ = registerTask(createFromIds, $C_{11}$, $C_{12}$, $C_{21}$, $C_{22}$)\tikzmark{right}$\ \ \, $\tikzmark{right2}
  \EndIf  \tikzmark{bottom}
  \Else
  \State $C$ = NIL  \tikzmark{fallbackmark}
  \EndIf
  \State \textbf{output:} $C$
\AddBraceNote{top}{bottom}{right}{execute}
\AddArrowNote{fallbackmark}{right}{right2}{fallback}
\end{algorithmic}
\caption{Pseudo-code for quadtree based matrix-matrix multiplication using the Chunks and Tasks programming model. \label{alg:pseudo_multiply}}
\end{algorithm}

\begin{algorithm}[h!]
\begin{algorithmic}[1]
  \State \textbf{input:} $A, B$
  \If{$A$ not NIL \textbf{and} $B$ not NIL}
  \If{lowest level}\tikzmark{add_top_e}
  \State $X$ = leafMatrixAdd($A$, $B$)
  \State $C$ = registerChunk($X$)
  \Else
  \For{$m = 1, 2$}
  \For{$n = 1, 2$}
  \State $C_{mn}$ = registerTask(add, $A_{mn}$, $B_{mn}$)
  \EndFor
  \EndFor
  \State $C$ = registerTask(createFromIds, $C_{11}$, $C_{12}$, $C_{21}$, $C_{22}$)\tikzmark{add_right}
  \EndIf \tikzmark{add_bottom_e}
  \Else
  \If{$A$ not NIL}\tikzmark{add_top_f}
  \State $C = A$
  \ElsIf{$B$ not NIL}
  \State $C = B$
  \Else
  \State $C$ = NIL
  \EndIf \tikzmark{add_bottom_f}
  \EndIf
  \State \textbf{output:} $C$
\AddBraceNote{add_top_e}{add_bottom_e}{add_right}{execute}%
\AddBraceNote{add_top_f}{add_bottom_f}{add_right}{fallback}%
\end{algorithmic}%
\caption{Pseudo-code for quadtree based matrix addition using the Chunks and Tasks programming model. \label{alg:pseudo_add}}
\end{algorithm}

\subsection{Chunk types for quadtree representation}\label{subsec:matrixchunk}

\begin{itemize}
\item[--]\emph{Matrix}: A basic matrix chunk type is used to represent
  nonzero submatrices in the quadtree representation. At each but the
  lowest level in the hierarchy, the matrix is divided into four
  submatrices represented by their chunk identifiers. At the lowest
  level, a leaf matrix type is used for matrix representation.
  Storage and addressing of zero submatrices is avoided at all levels in
  the hierarchy. Zero submatrices are represented by NIL chunk
  identifiers. Note that a NIL chunk identifier can appear at any level
  in the hierarchy.
  The matrix dimension is also stored along with the
  maximum allowed dimension for leaf matrices. This basic chunk type
  is the natural Chunks and Tasks implementation of a quadtree matrix
  representation as defined by Wise and Franco~\cite{WiseAndFranco1990}.
  When setting up the quadtree structure the matrix is split so that a
  predetermined uniform blocksize at each level in the hierarchy is
  achieved as far as possible. This ensures that when two matrices $A$
  and $B$ are combined in e.g. a multiplication or addition operation,
  the submatrix dimensions of $A$ and $B$ will match.  The blocksize
  used at leaf level should be chosen to achieve an appropriate
  granularity of chunks and tasks.  This is a trade-off: the leaf level
  chunks/tasks should be small enough to allow sufficient parallelism
  for the given computational resource.  At the same time, each leaf
  level chunk/task should contain enough data/work to make the
  administration overhead of the runtime library negligible. The matrix
  dimensions at higher levels in the hierarchy are directly determined
  by the lowest level blocksize, by multiplying by a factor of 2 for
  each level.
  A submatrix in
  the quadtree represented by a matrix chunk does not contain any
  global information such as the global matrix dimension or its
  location in the entire matrix, i.e.~row and column offsets.
\item[--]\emph{Matrix parameters}: A chunk type for matrix parameters
  is used to convey information needed in the construction of matrix
  chunks.  The chunk includes information about the matrix dimension
  and the leaf matrix dimension, and whenever needed information about
  the location of the matrix (its rowwise and columnwise offsets from
  the upper left corner) in the global matrix. It is also used to
  store information needed by the leaf matrix type.
\end{itemize}

\subsection{Task types for regular matrix-matrix multiplication}

\begin{itemize}
\item[--]\emph{$C = AB$, $C = A^TB$, $C = AB^T$, and $C = A^TB^T$}:
  The task types for regular and transposed matrix-matrix
  multiplication takes two matrix chunks, described above, and returns
  the product.  If the input matrices are both at the lowest level in
  the hierarchy, the corresponding leaf matrix multiplication is
  invoked.
  Otherwise matrix multiplication and matrix addition tasks for child
  submatrix multiplication and addition are registered for
  execution. The results are collected into the result matrix with a
  task for creation of a matrix chunk from four submatrix chunk
  identifiers. See the pseudo-code in
  Algorithm~\ref{alg:pseudo_multiply} for the $C = AB$ case.
\item[--]\emph{$C = A+B$}:
  The task type for matrix addition takes two matrix chunks and
  returns their sum. If the input matrices are at the lowest level,
  the addition in the leaf matrix library is performed.
  Otherwise, tasks for child submatrix addition are registered for
  execution, and the results are collected with a task for creation of
  a matrix chunk from child submatrix identifiers. See the pseudo-code
  in Algorithm~\ref{alg:pseudo_add}.
\item[--]\emph{Creation from submatrix identifiers}:
  A task type for creation of a matrix chunk from four submatrix chunk
  identifiers is needed since chunks are read-only after the point of
  registration. Since the submatrix chunk identifiers are included in
  the matrix chunk, it is not possible to construct the matrix chunk
  before the construction of the submatrices. Therefore, this task
  type is used whenever a matrix that depends on the results of other
  tasks needs to be constructed.
  This task type also takes a chunk of matrix parameters as input.
\end{itemize}

\subsection{Additional task types for symmetric matrix square}\label{subsec:sysq-task-types}
\begin{itemize}
\item[--]\emph{$C = A^2$ where $A$ is symmetric}:
A symmetric matrix square task squares a symmetric matrix in upper
triangular storage. At the lowest level the corresponding leaf matrix
symmetric matrix square operation is executed. At higher levels, the
symmetric matrix square task registers symmetric rank-k, symmetric
square, symmetric multiply, and matrix addition tasks to compute the
submatrices in the product matrix.  The results are collected into the
result matrix chunk with a task for creation of a matrix chunk from
submatrix identifiers. In general, a symmetric matrix square task
directly or indirectly makes use of all but the $C = A^TB^T$ task type
in this section.
Therefore, a benchmark of the symmetric matrix square operation covers
nearly all task types presented here.

\item[--]\emph{$C = AB$, where $A$ or $B$ is symmetric}:
Symmetric matrix multiply is the task type for multiplication of two
matrices where either the first or the second multiplicand is a
symmetric matrix in upper triangular storage.  At the lowest level the
corresponding multiplication operation for symmetric matrix multiply
is executed. At higher levels, tasks for regular matrix
multiplication, symmetric matrix multiply, and matrix addition are
registered and the results are collected into the result matrix chunk
with a task for creation from submatrix identifiers.

\item[--]\emph{$C = AA^T$ and $C = A^TA$}:
With this task, the so-called symmetric rank-k operation is performed;
a symmetric matrix in upper triangular storage is constructed from the
product of a general matrix and its transpose.  At the lowest level
the symmetric rank-k operation of the leaf matrix library is used. At
higher levels, symmetric rank-k, matrix multiplication, and matrix
addition tasks are registered and
a task for creation from submatrix identifiers is used
to collect the results into the result matrix
chunk, as for the other task types.

\end{itemize}

\section{Leaf matrix types}\label{sec:leaf_matrix_types}
 As discussed above, different leaf matrix representations may be used
 at the lowest level in the quadtree. 
A leaf matrix type used together with our chunk quadtree
representation has to implement some basic functionality such as
serialization routines. The leaf matrix type also has to implement
functionality needed by task types used together with the leaf matrix
type. When implementing a leaf matrix type, one can assume that the
matrix data fits in the memory of a single compute node. All
functionality in the class has to be thread-safe, since the Chunks and
Tasks library must be able to execute several leaf tasks and call the
serialization routines simultaneously.

\subsection{Block-sparse leaf matrix type}\label{subsec:blocksparse}

In the present work we are using a block-sparse leaf matrix type.
The block-sparse matrix class uses a uniform blocksize configurable
via the matrix parameters chunk type as discussed in
Section~\ref{subsec:matrixchunk}.
Submatrices are kept in a simple two-dimensional array where only
non-zero submatrices are allocated.
The
leaf matrix library makes use of the Basic Linear Algebra Subprograms
(BLAS) \cite{blas-level3} for submatrix-submatrix multiplications on CPUs
and the NVIDIA CUDA Basic Linear Algebra Subroutines (cuBLAS)
\cite{cublas} for submatrix-submatrix multiplications on 
graphics processing units (GPUs).
An
advantage of using the BLAS and cuBLAS library interfaces is that we
can take advantage of optimized BLAS and cuBLAS
library implementations.  We have for example observed substantial
performance improvements of the cuBLAS library when going from
Cuda~5.0 to Cuda~6.5 (see the caption of Table~\ref{tab:performance_batched_gemm}).

\begin{figure}
  \begin{center}
    \includegraphics[width=0.6\textwidth]{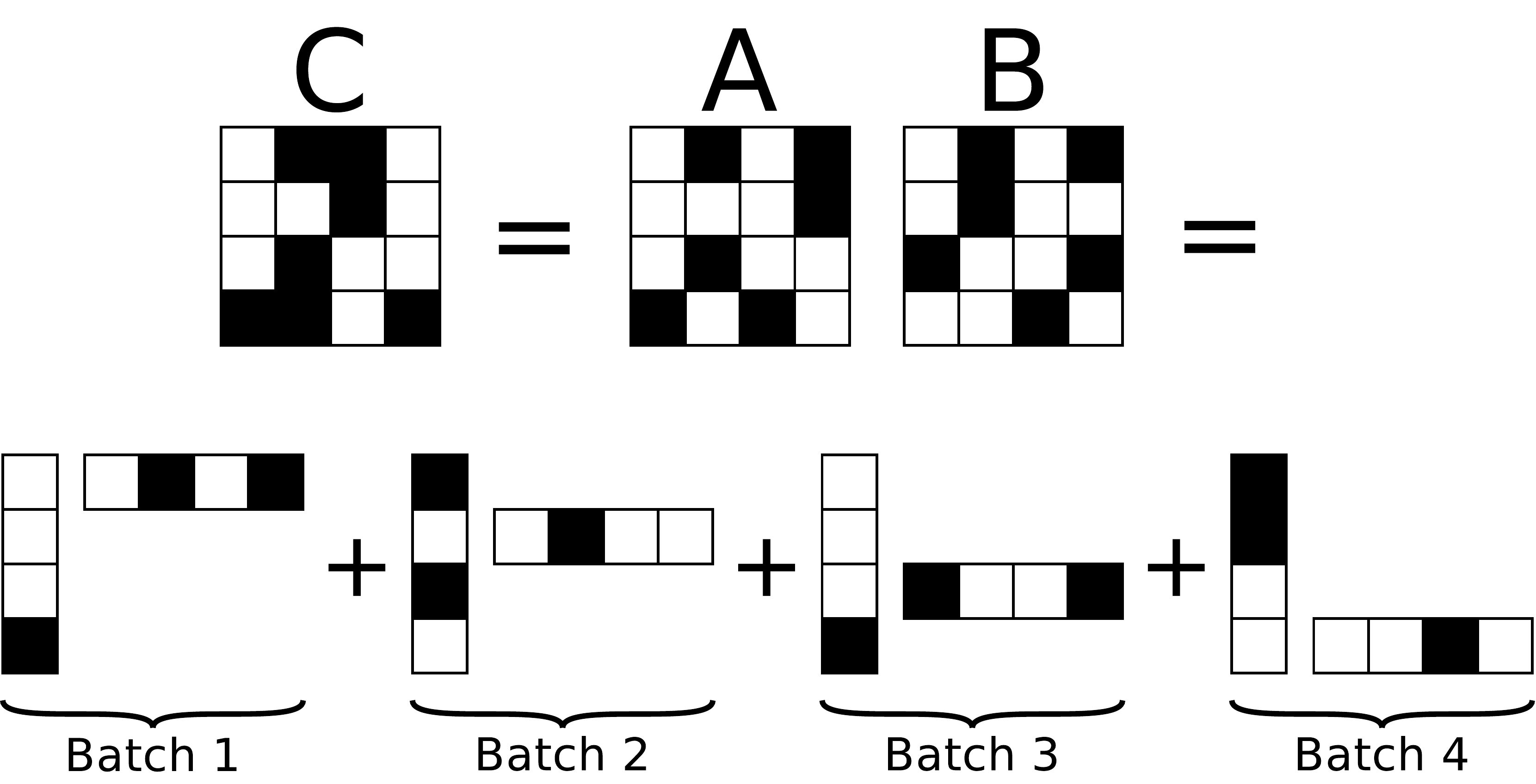}   
  \end{center}
  \caption{Illustration of a sparse matrix product as a sum of sparse
    outer products. All multiplies within a batch are independent
    making it possible to use the batched gemm API in
    cuBLAS. \label{fig:block-sparse-mmul}}
\end{figure}

Our block-sparse leaf matrix type targets problems where a block size
around 16-64 is appropriate. The regular gemm operation in cuBLAS is
inefficient for such small matrix dimensions. Therefore, we are
instead making use of the batched gemm API in cuBLAS. The routine
executes a batch of small matrix-matrix multiplications. The
operations in a batch should be independent in the sense that none of
the multiplications are allowed to write to the same product matrix.
The block-sparse multiplication can be expressed as a sum of outer
products, see Figure~\ref{fig:block-sparse-mmul}. Each outer product is
a batch of small matrix-matrix multiplications and all multiplies
within a batch are independent.

We would like to make use of both CPUs and GPUs, if any. If there is a
GPU available, the multiply will be processed by the GPU. Otherwise the
multiply will be processed by a CPU core. Since there will be other
threads that execute similar tasks, a GPU may become available during
the calculation. Then, it is generally good if the remaining work can be
offloaded to the GPU. To achieve this we are using
Algorithm~\ref{alg:process_batches}.  Using this algorithm, load balancing
between the CPUs and GPUs is achieved when several threads executing
leaf matrix multiplies are running on the host.  An alternative could
be to use a Chunks and Tasks library that let idle host threads
reexecute already running tasks whenever there is no more work, as in
\cite{beri2014scheduling}. The application programmer would then not
have to worry about feeding smaller portions to the CPU, and such an
approach could also help in case of various failures.  On the other
hand, some tasks would be executed more than once.

\begin{algorithm}
\begin{algorithmic}[1]
  \State Get list of batches (CPU) 
  \While{not done}
  \If{free GPU slot}
  \State Process remaining batches on GPU 
  \Else
  \State Process one batch on CPU 
  \EndIf
  \EndWhile
\end{algorithmic}
\caption{Algorithm for load balanced processing of batch
  lists.  \label{alg:process_batches}}
\end{algorithm}

\subsection{Efficient utilization of both GPUs and CPUs}

To make efficient use of both GPUs and CPUs on each node, data
transfers to/from GPUs should as far as possible be overlapped with
computation and the load should be shared between the GPUs and the CPU
cores. In addition, in case the amount of work in each task varies, it
is preferable to run small tasks on the CPU cores and let the GPUs
handle the computationally heavier tasks.

Overlap of data transfers and computation is achieved by keeping two
slots for each device (GPU), allowing one thread to do computations
while another thread is transferring data. In order to have a task
ready for execution on each device, a bounded queue is used, with a
queue length equal to the number of devices. To allow computationally
heavy tasks to run primarily on the GPUs, the queue is prioritized
according to a measure of the expected amount of work needed to
execute each task.  For tasks that do not fit in the queue, the work
is performed on a CPU core until a GPU becomes available, see
Algorithm~\ref{alg:process_batches}. This gives load balancing between
GPUs and CPUs provided that the number of threads is large enough; for
small numbers of threads only GPUs are used.

Since the preparatory work needed to get the list of batches for each
task is performed on a CPU core before checking for a place in the
queue, computation on the GPU is overlapped not only with data
transfers to/from the device memory but also with the preparatory
work.

Note that the Chunks and Tasks C++ interface presented in
\cite{chunks-and-tasks} does not include support for data transfers
between the host and GPU memory nor does it assist in scheduling of
tasks on GPUs, as does for example StarPU \cite{starpu}. The Chunks
and Tasks library is unaware of any devices that may be installed on
the compute nodes.
The code for deciding when to offload work to the GPU, as well as data
transfers to/from GPU memory, is part of the task execution. This
means that data that is used by several tasks running on the same GPU
will be transferred to the device memory once for each task.
However, as will be seen in Section~\ref{sec:gpu-calculations}, the
design described above achieves both hiding of data
transfer costs and sharing of the workload between GPUs and CPUs.

\section{Computational costs associated with the quadtree representation}\label{sec:quadtree-effects}

In this section, we first consider the number of tasks for different
sparsity patterns, and then use those results to get theoretical
estimates of computation and communication costs.

The total number of tasks includes both addition and multiplication
tasks. However, for the purpose of studying the scaling behavior of
the total number of tasks it is sufficient to consider only the number
of multiplication tasks since this is always larger than the number of
addition tasks. This can be understood in the following way. The input
submatrices to any addition task are temporaries that come from
previous addition or multiplication tasks. The input submatrices to
any multiplication task come from the original input matrices ($A$ and
$B$). The output submatrix resulting from any multiplication or
addition task can appear as input to at most one addition task.
Therefore, the tasks generating the trace of temporaries that precedes
any addition task form a binary tree in which every node has 0 or 2
children with multiplication tasks as leaves and addition tasks as
non-leaves.  Such a tree has more leaves than non-leaves.  Thus, the
total number of addition tasks is strictly bounded by the total number
of multiplication tasks.

\subsection{Total number of tasks}
We will here consider the total number of matrix-matrix multiplication
tasks for different sparsity patterns. Tasks executed at higher levels
can be seen as administration work required to determine which
low-level tasks are needed. A key issue is how much such extra
administration work that is generated due to the quadtree
representation.

In this section, we consider a quadtree representation with blocksize 1
at the lowest level. In practical calculations, a larger blocksize will
typically be used for performance reasons; we use blocksize 1 here in
order to more clearly see the effects of the quadtree structure. The
use of a larger blocksize will correspond to merging several of the
lowest levels, leaving it to the leaf matrix library to handle any sparsity there.

We first consider a case with little data locality, a random sparsity
pattern where the nonzero matrix elements are uniformly randomly
distributed. The probability $\delta$ to find a nonzero element at a given
position is the same everywhere in the matrix and uncorrelated to the
position of other nonzero matrix elements.

\begin{figure}
  \begin{center}
    \includegraphics[width=0.48\textwidth]{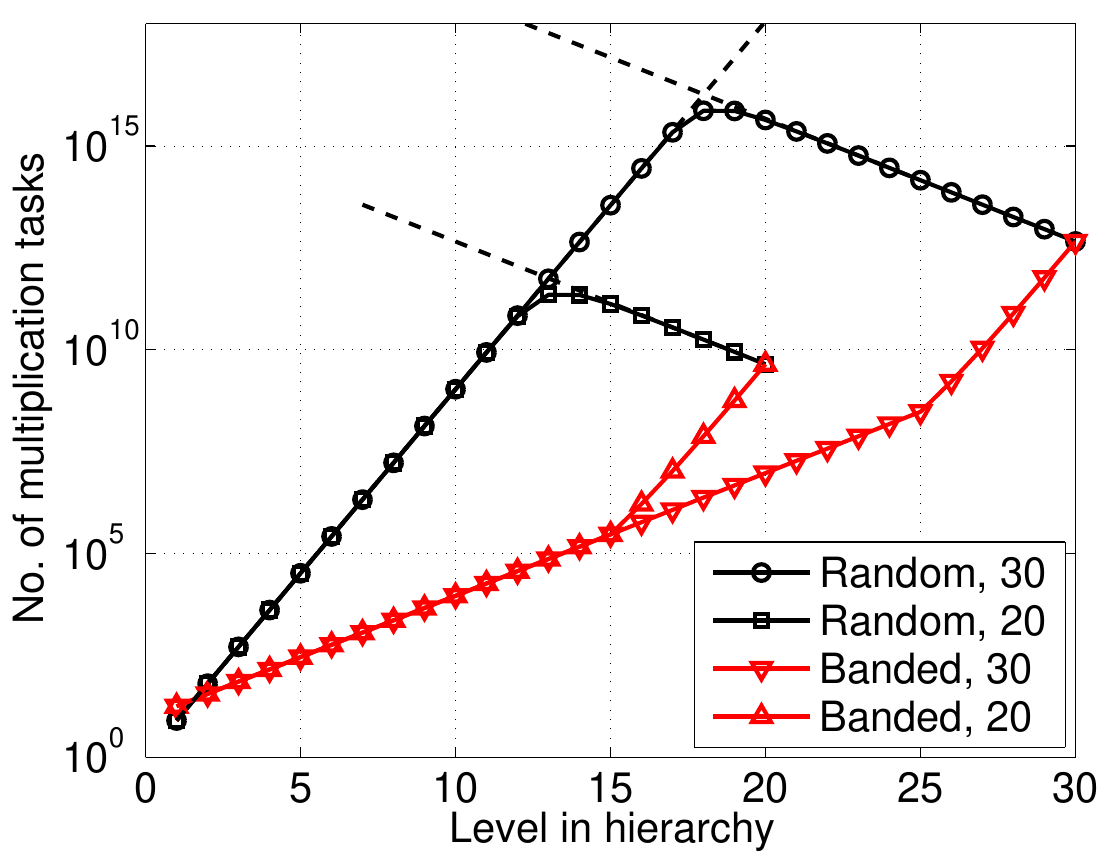}
  \end{center}
  \caption{Number of matrix-matrix multiplication tasks at different
    levels in the hierarchy for random and banded matrices with matrix
    dimension $N=2^L$ with $L=20$ and $L=30$.  The target number of
    nonzero elements per row is 65, corresponding to $\delta = 65/N$
    in the random case and $k=5$ in the banded case.  The dashed lines
    indicate the upper bounds given by \eqref{eq:bound_random_A} and
    \eqref{eq:bound_random_B}. \label{fig:ntasks_rand_and_banded}}
\end{figure}

Let the levels in the hierarchy be labeled such that level $l = 0$ is
the highest level (the root of the tree) and level $l = L$ is the
lowest (leaf) level.  Let $N$ be the matrix dimension and $N_l = 2^l$
such that $N_l^2$ is the total number of submatrices at level $l$ and
$N_L = N$.  If we denote the probability of a submatrix at level $l$
being nonzero as $\delta_l$, the expected number of matrix-matrix
multiplication tasks at level $l$ is
\begin{equation}\label{eq:no_of_mmul_tasks_at_level_random}
  C_l^{\mathrm{random}} = N_l^3 \delta_l^2 = 2^{3l} ( 1 - (1-\delta)^{n_l} )^2
\end{equation}
where $n_l = 2^{2(L-l)}$ is the total number of elements (including
both zeros and nonzeros) in each submatrix at level $l$.  The
relationship~(\ref{eq:no_of_mmul_tasks_at_level_random}) is
illustrated in Figure~\ref{fig:ntasks_rand_and_banded}.

Since $\delta_l \leq 1$, it follows that 
\begin{equation} \label{eq:bound_random_A}
  C_l^{\mathrm{random}} \leq N_l^3 = 8^l \quad \textrm{for all } l. 
\end{equation}
Furthermore, the
probability $\delta_l$ of a submatrix being nonzero satisfies the relation
$\delta_l = 1-(1-\delta_{l+1})^4 = 4\delta_{l+1}-6\delta_{l+1}^2+4\delta_{l+1}^3-\delta_{l+1}^4 \leq
4\delta_{l+1}$. Therefore, we also have that 
\begin{equation} \label{eq:bound_random_B}
  C_l^{\mathrm{random}} \leq N_l^3(4^{L-l}\delta)^2 = \frac{16^L\delta^2}{2^l} \quad \textrm{for all } l. 
\end{equation}
Although both inequalities \eqref{eq:bound_random_A} and
\eqref{eq:bound_random_B} are valid for all $l$,
\eqref{eq:bound_random_A} is tight for low levels while
\eqref{eq:bound_random_B} is tight for high levels, as seen in
Figure~\ref{fig:ntasks_rand_and_banded}.

Let $x = \log_2(N) + \frac{log_2(\delta)}{2}$ be the point where the two
bounds above intersect, given by solving $4^{L-x}\delta = 1$. Then, 
$8^x = (\delta N^2)^\frac{3}{2}$,
$8^{\lfloor x \rfloor} \leq 8^x$, and
$\frac{16^L\delta^2}{2^{\lfloor x \rfloor+1}} < \frac{16^L\delta^2}{2^{x}} =  8^x$. Therefore,
assuming $\delta \geq 1/N^2 \Longrightarrow x\geq 0$ the expected total number of tasks
\begin{align}
  \sum_{l = 0}^L C_l^{\mathrm{random}} & \leq \sum_{l = 0}^{\lfloor x \rfloor} 8^l + \sum_{l = {\lfloor x \rfloor}+1}^L \frac{16^L\delta^2}{2^l} \\
  & = 8^{\lfloor x \rfloor} \sum_{l = 0}^{\lfloor x \rfloor} \frac{1}{8^l} 
  + \frac{16^L\delta^2}{2^{\lfloor x \rfloor+1}} \sum_{l = 0}^{L-\lfloor x \rfloor-1} \frac{1}{2^l} \\
  & < 8^x \left( \frac{1}{1-1/8} + \frac{1}{1-1/2} \right) \\
  & = (3\tfrac{1}{7})(\delta N^2)^{3/2}.
\end{align}
We note that even though the number of tasks at leaf level is
$\mathcal{O}(N^3\delta^2)$, the total number of tasks is
$\mathcal{O}(N^3\delta^{3/2})$ due to excessive administration work at
higher levels.

As a simple case with data locality, we consider banded matrices with
bandwidth $b = 2d + 1$ where for simplicity we assume that $d = 2^k$
for some $k \geq 0$. In this case, the number of matrix-matrix
multiplication tasks at level $l$ is bounded by
\begin{equation}\label{eq:no_of_mmul_tasks_at_level_banded}
C_l^{\mathrm{banded}} < N_l b_l^2 = 2^l (2 d_l + 1)^2
\end{equation}
where
\begin{equation}
  d_l=\begin{cases}
  1          \quad \mathrm{for} \quad l < L-k, \\
  2^{l-(L-k)}  \quad \mathrm{for} \quad l \geq L-k.
  \end{cases}
\end{equation}

As seen in Figure~\ref{fig:ntasks_rand_and_banded}, the case with data
locality gives a very different behavior of the number of tasks on
each level; most of the work is concentrated at the lowest levels in
the hierarchy.
The total number of tasks is bounded by 
\begin{align}
  \sum_{l=0}^L C_l^{\mathrm{banded}} 
  & < \sum_{l=0}^{L-k-1} 2^l 3^2 + \sum_{l=L-k}^{L} 2^l(2 \cdot 2^{l-(L-k)}+1)^2 \\
  & < (4\tfrac{4}{7}d^2+5\tfrac{1}{3}d+2+\frac{9}{d})N.
\end{align}
We note that the total number of tasks is proportional to the number
of tasks at the lowest level, i.e. no excessive administration work is
going on at higher levels.

\begin{figure}
  \begin{center}
    \includegraphics[width=0.48\textwidth]{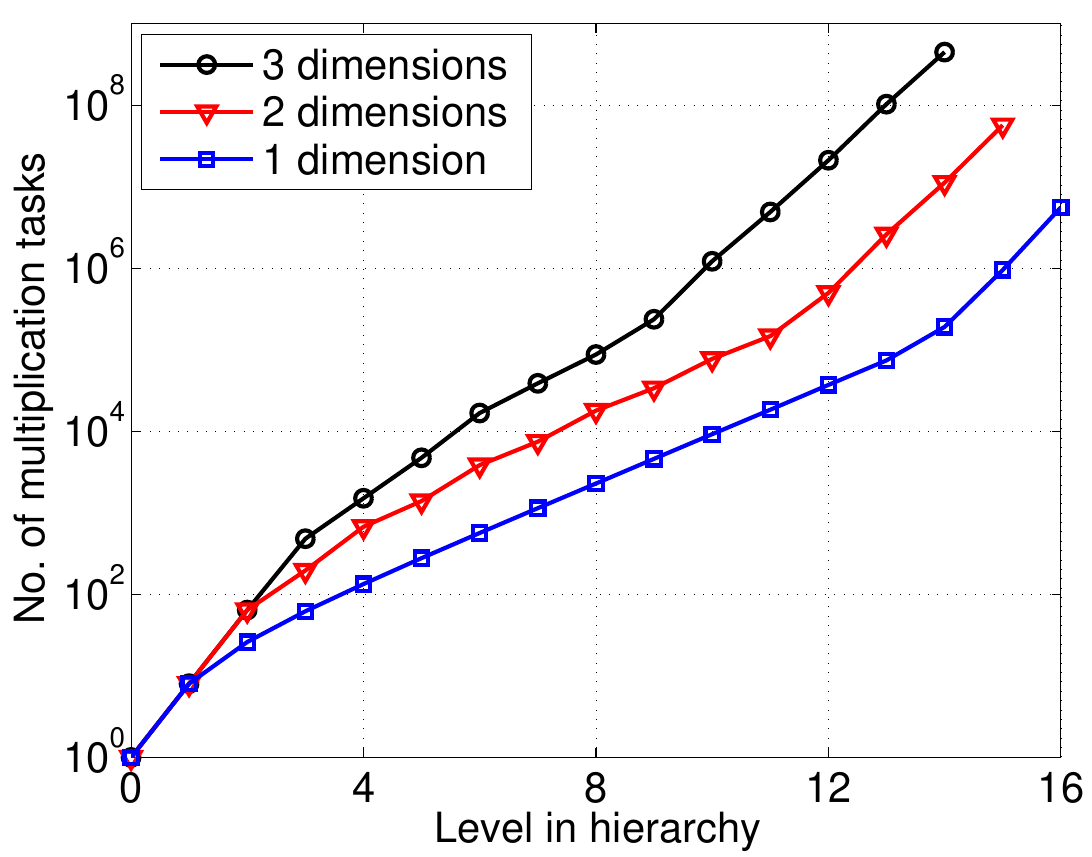}
    \includegraphics[width=0.48\textwidth]{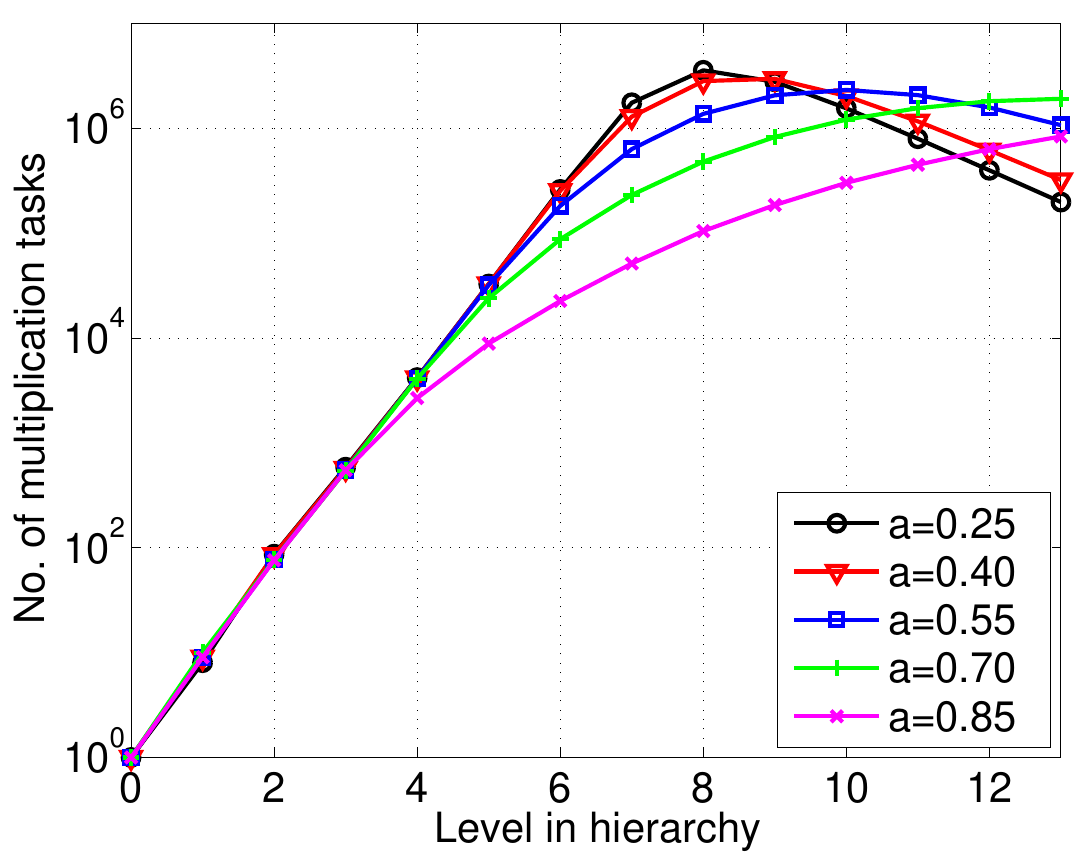}
  \end{center}
  \caption{Number of matrix-matrix multiplication tasks for matrices
    with different sparsity structures. Left: overlap matrices for
    artificially constructed 1d, 2d, and 3d molecules. Right: R-MAT
    graph matrices with parameters corresponding to different degrees
    of data locality. See the text for
    details. \label{fig:ntasks_1d_2d_3d_and_rmat}}
\end{figure}

As an example of sparsity structures appearing in physical
applications, we consider overlap matrices for systems of evenly
distributed particles in $D \geq 1$ spatial dimensions with one
spherically symmetric basis function per particle, ordered using a
recursive divide-space procedure. We consider finite systems (no
periodicity). A
matrix element $A_{ij}$ is nonzero if the distance between particles
 $i$ and $j$ is smaller than some radius $R$. Note that this
is a kind of sparsity structure found in many applications in physics
and chemistry, where each matrix element is often related to a pair of
particles or other objects in a physical system; sparsity arises from
the fact that only matrix elements that correspond to objects that are
sufficiently close in space are nonzero.
In $D$ spatial dimensions with the chosen ordering of basis functions,
the $N_l$ blocks at level $l$ can be seen as corresponding to a set of
$N_l$ spatial boxes such that all basis functions (or particles) in a
given block are contained in the corresponding box. The number of
multiplication tasks at a given level can be estimated by
\begin{equation}\label{eq:no_of_mmul_tasks_at_level_overlap}
C_l^{\mathrm{overlap}} < N_l M_l^2 = 2^l M_l^2
\end{equation}
where $M_l$ is the number of spatial boxes that can be reached by a
sphere of radius $R$. For high levels where the width of spatial boxes
is larger than $R$, $M_l$ is determined by the number of neighboring
boxes, $M_l = 3^D$, independently of $l$. For lower levels $M_l$ is
proportional to the volume of a $D$-dimensional sphere of radius
$\frac{R}{h_l}$ where the width of boxes $h_l \propto
2^{\frac{L-l}{D}}$, giving $M_l \propto R^D 2^{l-L}$. Therefore,
analogously to the banded matrix case, for high levels
$C_l^{\mathrm{overlap}} \propto 3^{2D} 2^l$ and for lower levels
$C_l^{\mathrm{overlap}} \propto R^{2D} 2^{3l-2L}$.
We note also that $C_l^{\mathrm{overlap}} \geq 2
C_{l-1}^{\mathrm{overlap}}$ for all $l \geq 1$.
Therefore, as for the banded matrix
case, the total number of tasks is proportional to the number of tasks
at the lowest level.
Numerical experiments for $D = 1, 2, 3$ are shown in the left panel of
Figure~\ref{fig:ntasks_1d_2d_3d_and_rmat}.
The test matrices were created using the {\sc Ergo} program \cite{linmemDFT} to compute
overlap matrices for artificially generated 1d, 2d, and 3d molecules
with one basis function per atom from the standard Gaussian basis set STO-3G, applying the default
recursive divide-space procedure to order the atoms. The molecules
were generated by placing hydrogen atoms on a $D$-dimensional grid with
separation 2~{\AA} and a uniform random displacement of up to $\pm 1$~{\AA} in each coordinate direction. The matrix size
was $2^{16} = 65536$ for $D = 1, 2$ and $40^3 = 64000$ for $D =
3$. Blocksize 1 was used for the 1d case while the 2d and 3d cases
used blocksize 2 and 4, reducing the number of hierarchy levels for those
cases by 1 and 2, respectively.

As an example where different degrees of data locality can be easily
investigated we consider multiplication of graph matrices constructed
using the R-MAT model~\cite{R-MAT}. We choose the R-MAT parameters
such that $b=c=d=\frac{1-a}{3}$ and perform tests for different values
of $a$ in the range $0.25 \leq a < 1$. Setting $a = 0.25$ essentially
corresponds to the random case described above, while increasing $a$
corresponds to increasing the data locality.  We use matrix dimension
$2^{13}=8192$ and a number of graph edges corresponding to 5 nonzero
elements per row, although for large $a$ values matrices become more
sparse due to multiple graph edges between the same nodes. The right
panel of Figure~\ref{fig:ntasks_1d_2d_3d_and_rmat} shows the number of
matrix-matrix multiplication tasks for a set of different $a$ values.
We note that most of the work is pushed towards lower levels as the
degree of data locality increases.

\subsection{Computation and communication costs} \label{subsec:CompAndCommCosts}

We will here assume that the tasks are as far as possible evenly
distributed over the computational nodes, i.e. that the total
execution time is given by $\mathcal{O}(T_1/p + T_\infty)$ where $T_1$
and $T_\infty$ are the serial and critical path execution times,
respectively and $p$ is the number of processes.  Such load balancing
can for example be achieved by work stealing
\cite{BlumofeAndLeiserson1999}.  
While the total number of addition tasks is always smaller than the
total number of multiplication tasks, the number of tasks along the
critical path is up to $\mathcal{O}((\log(N))^2)$ for the additions
and $\mathcal{O}(\log(N))$ for the multiplications.  Therefore, we
take $T_1$ to be proportional to the total number of multiplication
tasks and $T_\infty = \mathcal{O}((\log(N))^2)$ given by the worst
case critical path length. Thus, for the random case we have that the
execution time is 
\begin{equation} 
  \mathcal{O}\left(\frac{(\delta N^2)^{3/2}}{p} + (\log(N))^2\right)
\end{equation} 
and for the banded case we have that the execution time is
\begin{equation}\label{eq:exec_time_banded}
  \mathcal{O}\left(\frac{d^2N}{p} + (\log(N))^2\right).
\end{equation}

We are interested in the communication costs in the weak and strong
scaling limits. We consider first a weak scaling test constructed by
keeping the number of nonzero matrix elements per row fixed and
increasing the matrix dimension together with the number of processes.
In the random case, this means that $\delta \propto \frac{1}{N}$ and
the number of leaf level tasks is $\mathcal{O}(N)$ but the total
number of tasks is $\mathcal{O}(N\sqrt{N})$. Assuming that all data
for each task needs to be communicated this means that each process
needs to receive data scaling as $\mathcal{O}(\sqrt{p})$ with the
number of processes. In the banded and overlap cases both the leaf
level and total number of tasks is $\mathcal{O}(N)$ and the average
amount of data received per process is $\mathcal{O}(1)$. Since these
results are based on the number tasks, the latency cost behaves in the
same way; the number of messages exchanged per process is proportional
to the number of tasks per process.
For strong scaling the number of tasks is constant, so the average
number of tasks per process and thereby also the average amount of
data communicated per process scales as $\mathcal{O}(1/p)$ for all
sparsity structures.

The above results for the quadtree representation can be compared to
the approach where a random permutation is employed to destroy data
locality followed by application of the Sparse SUMMA
algorithm~\cite{SparseSUMMA2008}.  Assuming that the random
permutation succeeds to evenly distribute the nonzero matrix elements,
the number of matrix elements that each process needs to fetch from
other processes becomes (see e.g.~equation (3.1) in
\cite{BulucGilbert2012})
\begin{equation} \label{eq:sqrtp_equation}
  \frac{2mN}{\sqrt{p}} 
\end{equation}
where $m$ is the number of nonzeros per row. Similarly to the above, a
weak scaling test can be constructed by keeping $m$ fixed and letting
$N \propto p$, leading to each process receiving data scaling as
$\mathcal{O}(\sqrt{p})$ with the number of processes. The weak and
strong scaling results are summarized in
Table~\ref{tbl:weak_and_strong_scaling}.

\begin{table}
\begin{center}
\begin{tabular}{lcc}
  \hline
  & Weak & Strong \\
  \hline
  \hline
  Quadtree - random  & $\mathcal{O}(\sqrt{p})$   & $\mathcal{O}(1/p)$ \\
  Quadtree - banded  & $\mathcal{O}(1)$          & $\mathcal{O}(1/p)$ \\
  Quadtree - overlap & $\mathcal{O}(1)$          & $\mathcal{O}(1/p)$ \\
  \hline
  SpSUMMA &  $\mathcal{O}(\sqrt{p})$ & $\mathcal{O}(1/\sqrt{p})$\\
  \hline
\end{tabular}
\end{center}
\caption{Communication costs. Scaling of the average amount of data received
  by each process with the number of processes $p$ in the weak and
  strong scaling limits for matrices with different sparsity
  patterns.  \label{tbl:weak_and_strong_scaling}}
\end{table}

\section{Performance evaluation}\label{sec:performance}
In this section, we will examine the performance of our block-sparse
matrix-matrix multiplication presented in Section~\ref{sec:chtml} when
linked to the publicly available Chunks and Tasks library
implementation CHT-MPI described in Section~\ref{subsec:cht-lib-impl},
using the block-sparse leaf matrix library of
Section~\ref{sec:leaf_matrix_types}.  Hereinafter, the number of
floating point operations for multiplication of two dense matrices
with dimension $N$ is counted as $2N^3$.

\subsection{Calculations on cluster of GPU-equipped nodes}\label{sec:gpu-calculations}
We will first present calculations performed on the Erik cluster at
the Lunarc computer center, Lund University, using CHT-MPI~1.1 compiled with Open~MPI~1.6.5 
and gcc~4.8.1, Cuda 6.5 and the Intel Math Kernel Library (MKL) version~11.1.
The Erik cluster consists of 24 nodes each with dual 64-bit, 8-core
Intel Xeon E5-2650 2.00 GHz processors.  The nodes are interconnected
with FDR InfiniBand and equipped with Nvidia Tesla K20m GPU cards: 16
nodes with 2 cards, 7 nodes with 4 cards, and 1 node with 8 cards.
Leaf matrix submatrix operations were performed with the MKL
implementation of BLAS on CPUs and with cuBLAS on the GPUs, as
described in Section~\ref{subsec:blocksparse}.

\begin{table}
  \begin{center}
    \begin{tabular}{l|cccccc}
      Matrix size      &    16 &    32 &    48 &    64 &    80 &    96 \\
      \hline
      Gflop/s (single) & 243.1 & 392.3 & 276.6 & 558.4 & 401.9 & 628.9 \\
      Gflop/s (double) & 147.7 & 210.5 & 231.3 & 244.0 & 263.8 & 270.5
    \end{tabular}
  \end{center}
  \caption{Practical peak performance figures for cuBLAS batched
    matrix-matrix multiplication in Cuda 6.5 on Nvidia K20. Computed
    from batches with 64000 matrix-matrix multiplications, $C_i =
    \beta C_i + \alpha A B, \ i = 1,2,\dots, 64000$ with $A$, $B$,
    and $C_i$ being $b \times b$ matrices with $b =
    16,32,48,64,80,96$.  The number of floating point operations is
    counted as $64000 \times 2b^3$. We note that the figures are up to
    40\% larger than what we obtained with the benchmark program
    provided by Nvidia. This is mainly due to the reuse of the $A$ and
    $B$ matrices for all 64000 multiplies in the present
    benchmark. Also, in the present benchmark timers on the GPU were
    used instead of timers on the CPU combined with
    synchronization. This results in larger Gflop/s values.
    Furthermore, switching from Cuda 5.0 to Cuda 6.5 gave performance
    improvements ranging from 20~\% for the ``double precision, block
    size 16'' case to 310~\% for the ``single precision, block size
    96'' case, for calculations that were otherwise identical.
    \label{tab:performance_batched_gemm}}
\end{table}

\begin{figure}
  \begin{center}
    \includegraphics[width=0.48\textwidth]{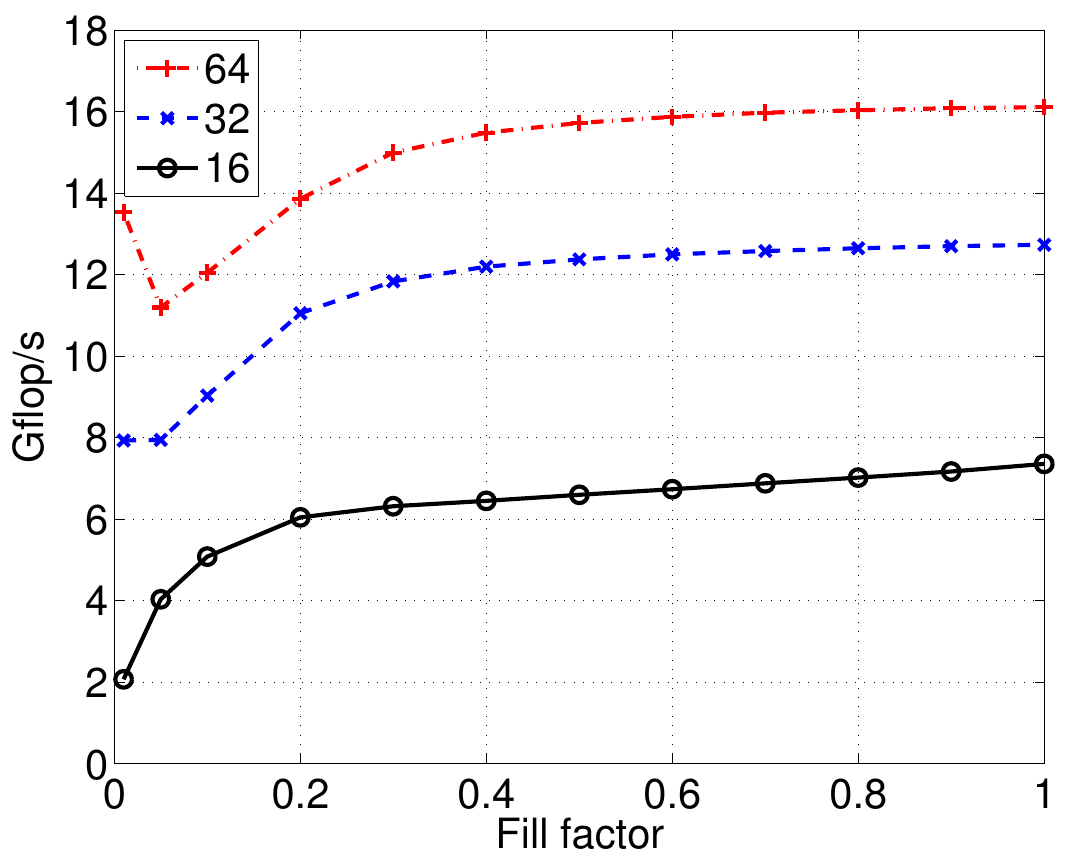}
    \includegraphics[width=0.48\textwidth]{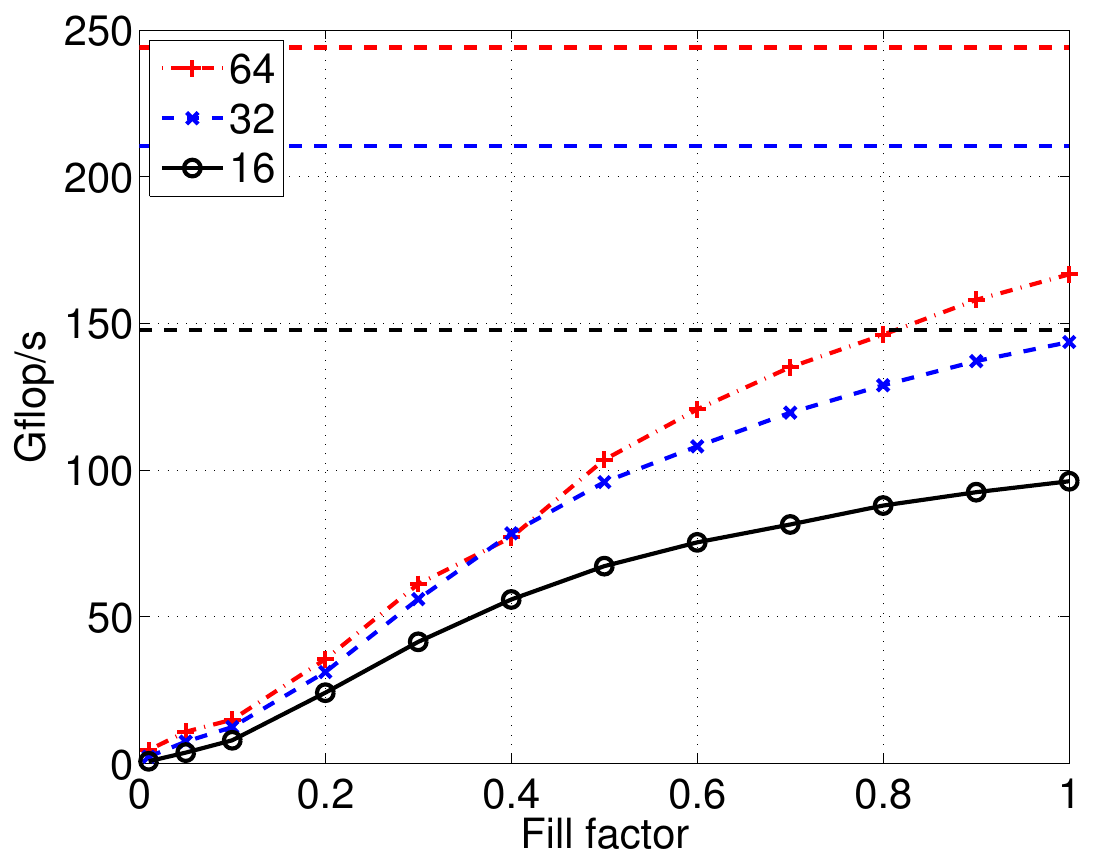}
  \end{center}
  \caption{Results of leaf block-sparse matrix library matrix-matrix
    multiplication test runs with double precision and matrix
    dimension $4096 \times 4096$, varying sparsity (fill factor), and
    blocksizes 16, 32, and 64 on the Erik cluster. The nonzero
    submatrices are randomly uniformly distributed over the matrix.
    Left: running on one of the CPU cores. Right: running on one of
    the CPU cores but processing the list of batches on one of the
    GPUs.  The dashed lines are practical peak performance figures
    computed from batches with 64000 $b\times b$ multiplies with $b =
    16,32,64$, not including any data transfers, see
    Table~\ref{tab:performance_batched_gemm}. \label{fig:leafmat_bench_double}}
\end{figure}

\begin{figure}
  \begin{center}
    \includegraphics[width=0.48\textwidth]{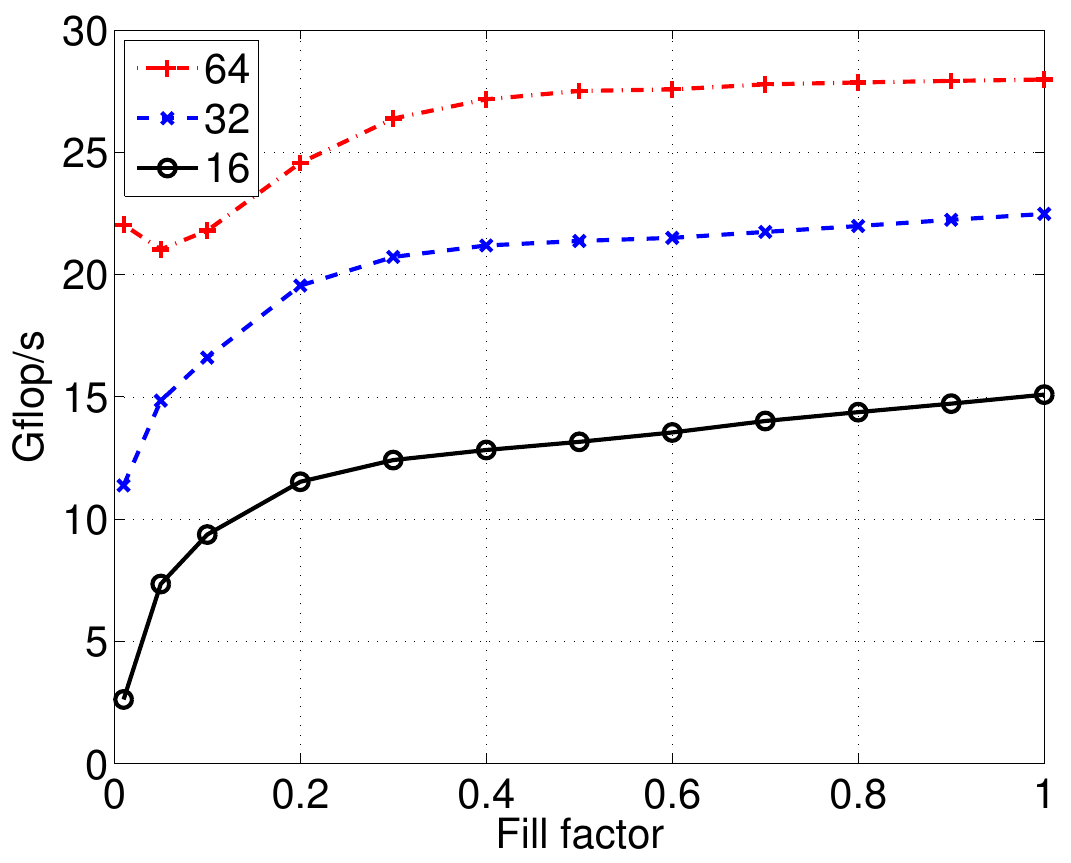}
    \includegraphics[width=0.48\textwidth]{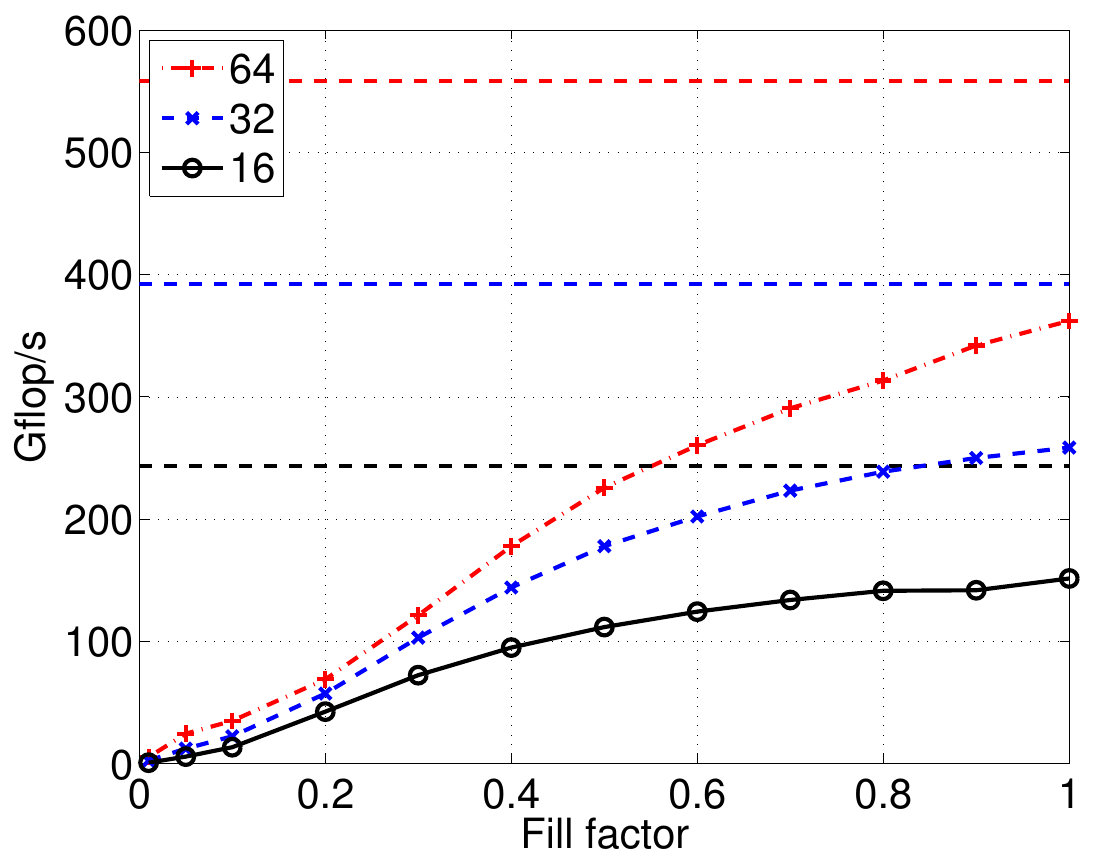}
  \end{center}
  \caption{Test runs for single precision corresponding to the double
    precision results in Figure~\ref{fig:leafmat_bench_double}. See
    that figure caption for more information.
 \label{fig:leafmat_bench_single}}
\end{figure}

Our first benchmark measures only the performance of the block-sparse
matrix-matrix multiplication used for the leaf multiplications in the
Chunks and Tasks matrix library. Thus, Chunks and Tasks is not
involved in this benchmark. We perform multiplications with matrix
dimension $4096 \times 4096$ with varying degree of matrix
sparsity. The nonzero submatrix blocks are randomly uniformly
distributed over the matrix to get a predetermined fill factor which
is the fraction of nonzero matrix elements. Results for double and
single precision are shown in Figures~\ref{fig:leafmat_bench_double}
and~\ref{fig:leafmat_bench_single}, respectively.  Practical peak
performance values (dashed lines) were calculated with a separate
benchmark program measuring the performance of the cuBLAS batched
matrix-matrix multiplication, see
Table~\ref{tab:performance_batched_gemm}. We are not quite reaching up
to those peak figures. The main reason is that the benchmark of the
leaf matrix library includes data transfers to/from the GPU which are
not included in the peak performance figures. The work needed to
prepare the list of batches is also included in the measured wall
time. However, when the leaf matrix library is used within the Chunks
and Tasks matrix library, there will be several threads that need to
execute leaf matrix multiplications. This means that both
communication to/from the GPU and preparation of batch lists on the
CPU can then be overlapped with computation on the GPU as described in
Section~\ref{sec:leaf_matrix_types}.


\begin{figure}
  \begin{center}
    \includegraphics[width=0.48\textwidth]{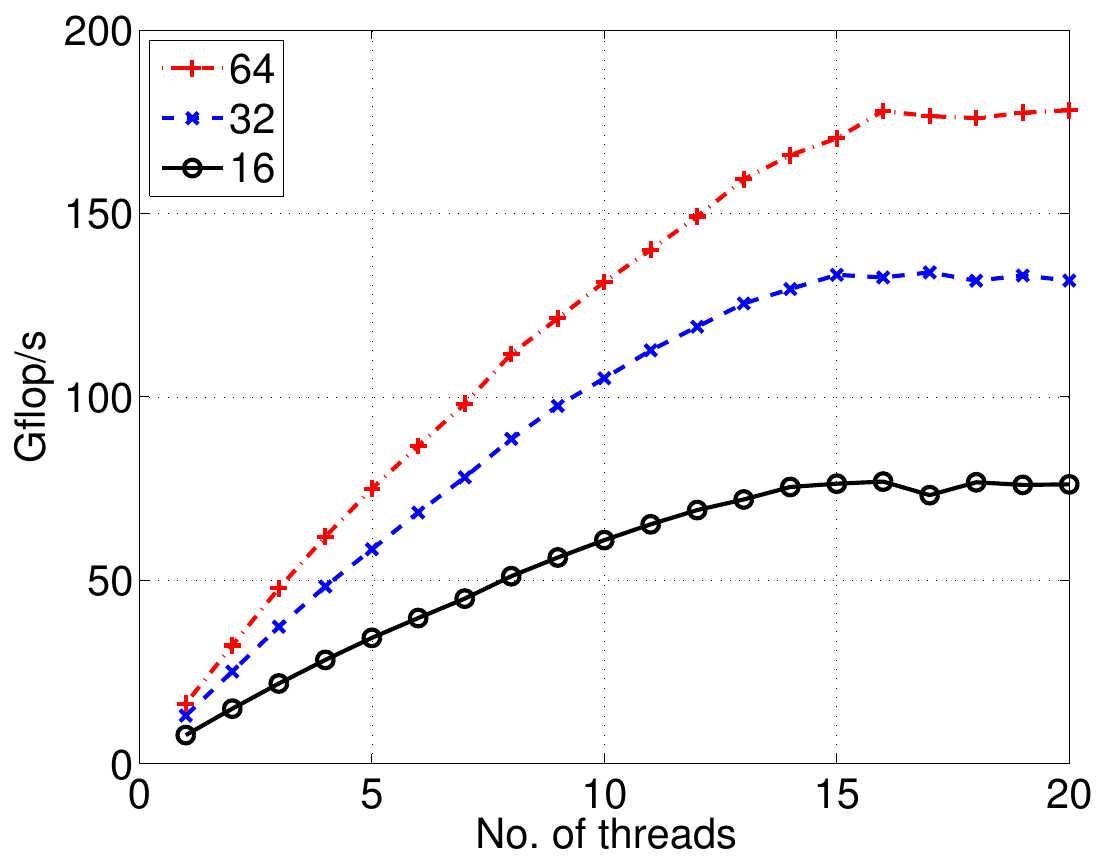}
    \includegraphics[width=0.48\textwidth]{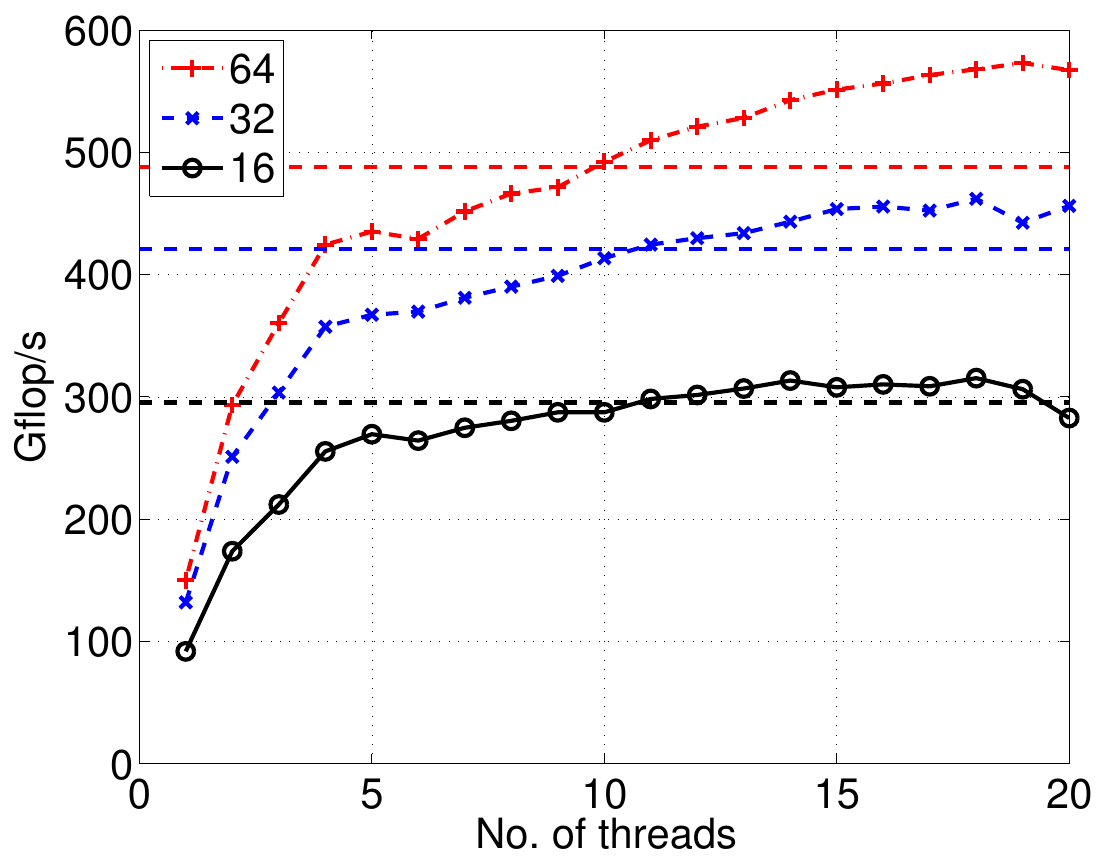}
  \end{center}
  \caption{Results of matrix-matrix multiplication test runs in double
    precision using the Chunks and Tasks matrix library for dense $25
    000\times 25 000$ matrices and blocksizes 16, 32, and 64 on a
    single Erik node equipped with two GPUs. A varying number of
    threads was used by CHT-MPI to execute tasks. The leaf matrix
    dimension was fixed to $4096 \times 4096$.  Left: running on the
    CPU cores. Right: running on both the CPU cores and the GPUs. The
    dashed lines indicate practical peak performance figures for the
    two GPUs, see Table~\ref{tab:performance_batched_gemm}. Note that
    up to 6 threads all batch lists are executed on the GPUs, see the
    discussion in the text. \label{fig:single_node_bench_double}}
\end{figure}

\begin{figure}
  \begin{center}
    \includegraphics[width=0.48\textwidth]{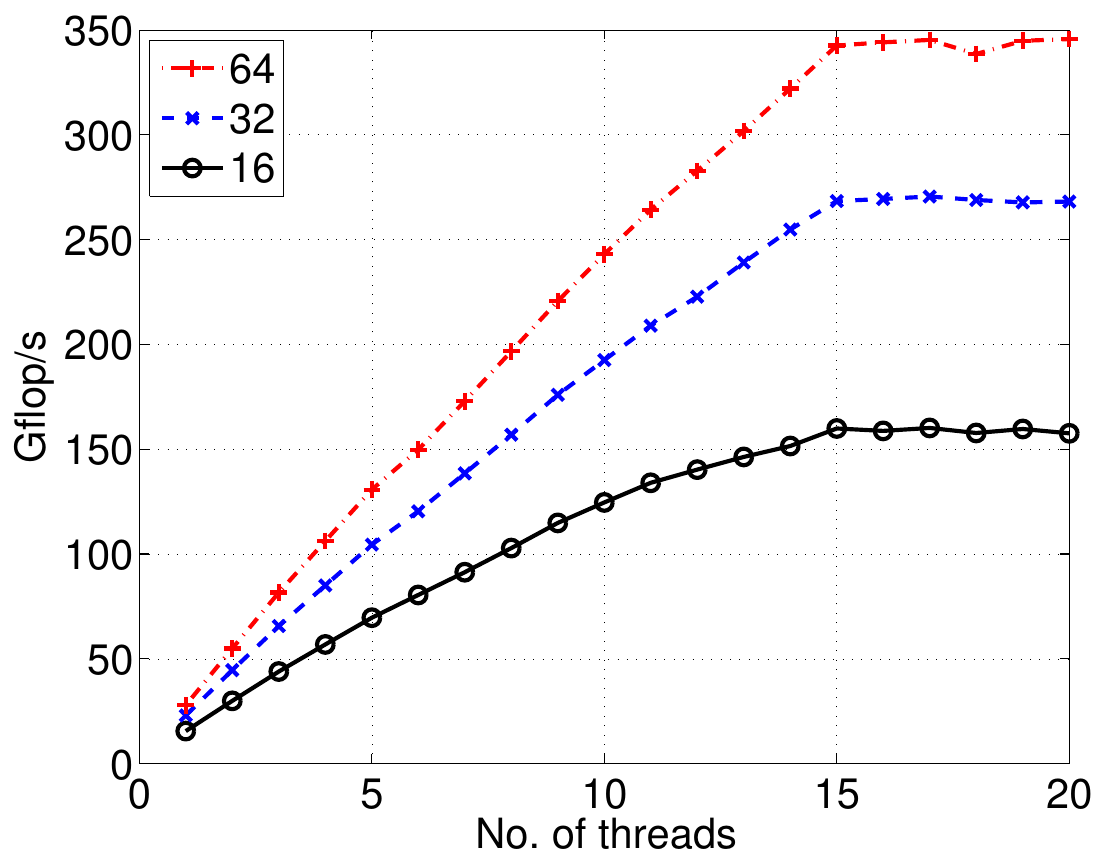}
    \includegraphics[width=0.48\textwidth]{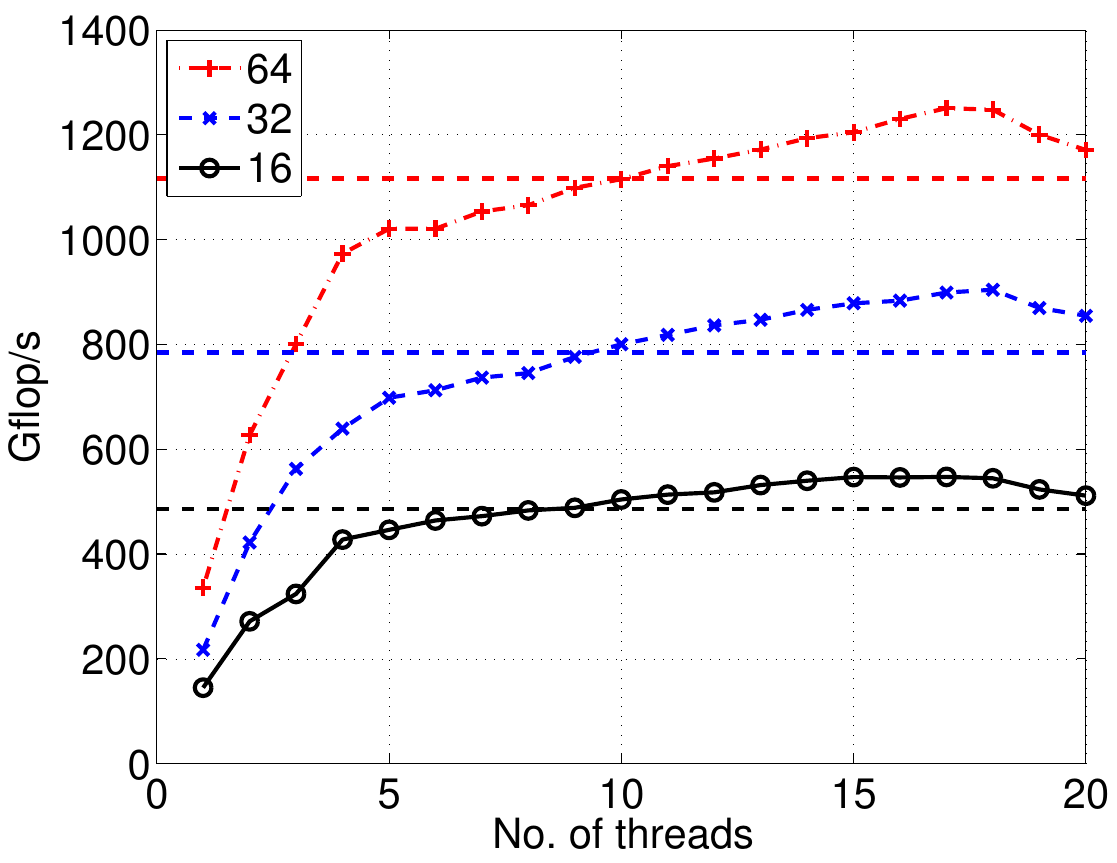}
  \end{center}
  \caption{Test runs for single precision and matrix dimension $40
    000\times 40 000$, otherwise corresponding to the double precision
    results in Figure~\ref{fig:single_node_bench_double}. See that
    figure caption for more
    information. \label{fig:single_node_bench_single}}
\end{figure}

This brings us to the next two benchmark figures where the Chunks and
Tasks library is used but only on a single computational node, see
Figures~\ref{fig:single_node_bench_double}
and~\ref{fig:single_node_bench_single}. The computational node is
equipped with 16 CPU cores and 2 GPUs.  Considering that the bounded
priority queue has length 2 and that there are 2 slots per GPU, this
means that up to 6 threads, no processing of batch lists will ever
occur on the CPU cores.  When 2 threads are executing tasks, there is
one GPU dedicated to each thread, similarly to the previous benchmark
figures.
Therefore, the performance improvement when going from 2 to 6 threads
comes solely from overlapping data transfers to/from the GPUs and
preparation of batch lists on the CPU with the processing of batch
lists on the GPUs. Comparing the practical peak performance limits to
the performance for 6 threads in
Figures~\ref{fig:single_node_bench_double}
and~\ref{fig:single_node_bench_single} shows that the performance for
6 threads reaches between 88\% and 95\% of the practical peak
performance values.
When increasing the number of threads
further, batch lists will also be processed on the CPU cores,
according to Algorithm~\ref{alg:process_batches}.  The figures show
that we are able to take advantage of both the CPU and GPUs in a load
balanced manner. We also note that no parametric models for task
execution times on different hardware were required, the load
balancing was achieved automatically without any information about the
computational power of the devices, other than the assumption that
a GPU is much more powerful than a CPU core. Block-sparse
matrix-matrix multiplication using GPUs has also been implemented in
the distributed block-compressed sparse row library, but using custom
computational kernels rather than the batched kernels in
cuBLAS~\cite{Borstnik2014}. The performance results in
Figure~\ref{fig:single_node_bench_double} are comparable to those in
Figure~8 in~\cite{Borstnik2014}.

\begin{figure}
  \begin{center}
    \includegraphics[width=0.48\textwidth]{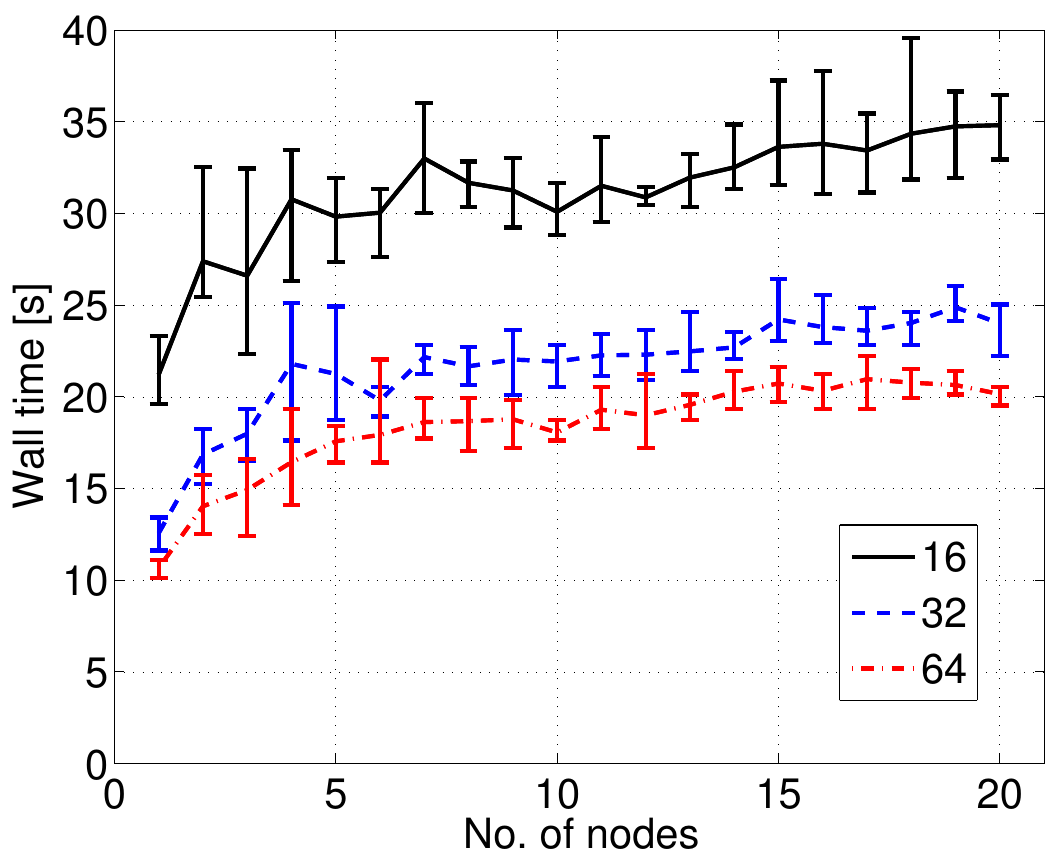}
    \includegraphics[width=0.48\textwidth]{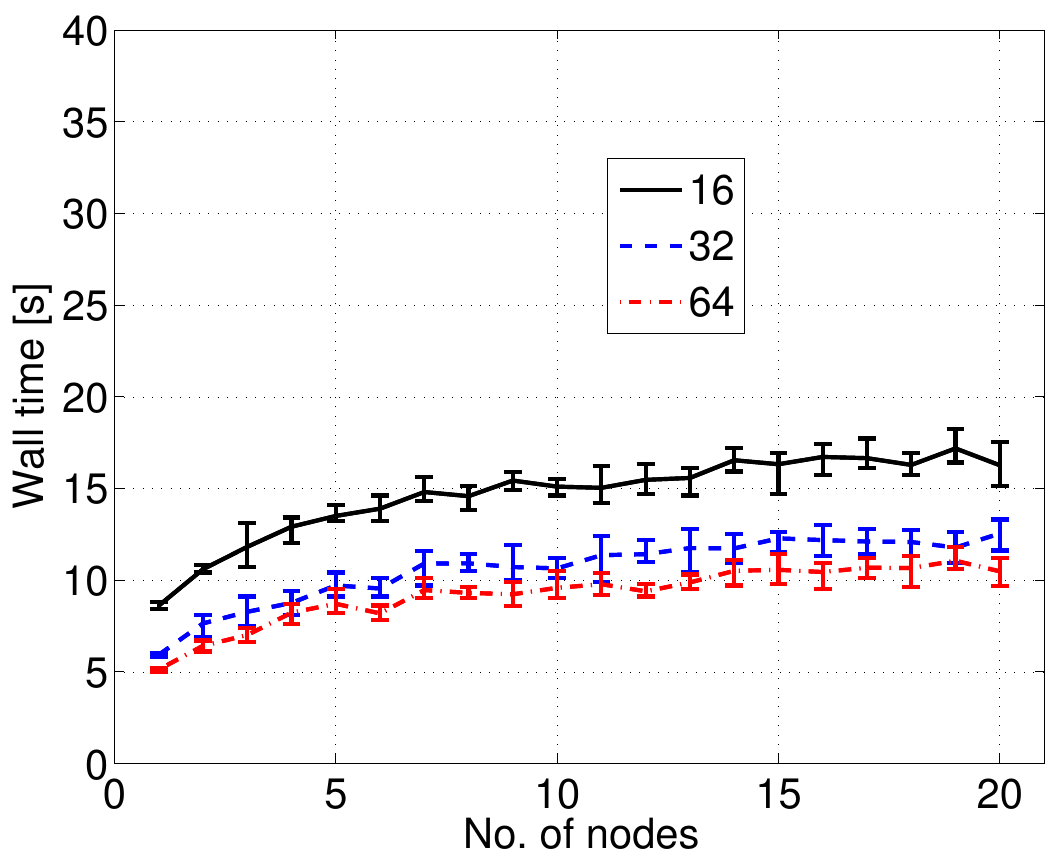}
  \end{center}
  \caption{Weak scaling test for banded matrices with bandwidth $2
    \times 4000+1$ and matrix dimension $40000n \times 40000n$, where
    $n$ is the number of nodes.  The calculations were performed in
    double precision with leaf matrix dimension $4096 \times 4096$ for
    internal leaf matrix block sizes 16, 32, and 64.  Left: Regular
    matrix-matrix multiplication. Right: Symmetric matrix square
    taking advantage of that the product matrix is symmetric.  For
    each case (no. of nodes and blocksize), the benchmark calculation
    was repeated 6 times, and we plot the smallest interval containing
    all 6 wall times.  Lines are drawn through the average
    wall time of the 6 benchmark calculations.  \label{fig:weak_scaling_erik}}
\end{figure}

In Figure~\ref{fig:weak_scaling_erik} we investigate the weak scaling
of our block-sparse matrix-matrix multiplication for a set of banded matrices with
fixed bandwidth but a matrix dimension that is increasing proportionally to the
number of computational nodes.
Each node used 17 worker threads and a chunk cache size of 5 GB.
In this case, we noticed fluctuations
in the execution time and have therefore carried out 6 test runs for
each case (no.~of nodes and blocksize). 
Note that according to the theoretical weak scaling results of
Section~\ref{sec:quadtree-effects}, we would expect at worst $\mathcal{O}((\log
(n))^2)$ scaling of the wall time with increasing number of nodes $n$.
Figure~\ref{fig:weak_scaling_erik} shows results for both regular matrix-matrix multiplication and
the symmetric matrix square
operation that assumes upper triangular storage of a symmetric matrix
and only computes the upper triangle of the symmetric product. The
implementation of
the symmetric matrix square
operation is straightforward using Chunks and
Tasks since all decisions regarding distribution of work and data are
handled by the runtime library.  The expected speedup of 2
compared to the regular multiplication is achieved.

The effective performance in Gflop/s for the tests in
Figure~\ref{fig:weak_scaling_erik} can be computed based on the number
of scalar multiplications and additions for multiplication of banded
matrices with bandwidth $2d+1$:
\begin{equation}
  2(N(2d+1)^2 - (5/3)d(d+1)(2d+1)).
\end{equation}
For the 1-node case in the left panel of
Figure~\ref{fig:weak_scaling_erik} this gives 222, 374, and 441
Gflop/s for block sizes 16, 32, and 64, respectively. For the 20-node
case the corresponding numbers are 2933, 4258, and 5065 Gflop/s.
These figures can be directly compared to the performance results in
the right panel of Figure~\ref{fig:single_node_bench_double}. In
particular, one can compare to the highest performance for each block
size.  This shows that the banded matrix test runs in
Figure~\ref{fig:weak_scaling_erik} retained 70\%, 81\%, and 77\% of
the performance for the 1-node case for block sizes 16, 32, and 64,
respectively. For the 20-node case the corresponding numbers are 47\%,
46\%, and 44\% compared to a perfect weak scaling scenario.  Since the
weak scaling efficiency per node has essentially leveled out at 20
nodes, similar efficiency can be expected also for larger
calculations.

\subsection{Application to overlap matrix in electronic structure program}\label{sec:smat-tests}

To test the applicability of our block-sparse matrix-matrix
multiplication in large-scale electronic structure calculations, we
have adapted parts of the linear scaling electronic structure code
{\sc Ergo} \cite{linmemDFT} so that the overlap matrix can be
constructed in parallel using Chunks and Tasks. This allows us to test
the symmetric matrix square operation for the overlap matrix. See the
description of the symmetric matrix square task type in
Section~\ref{subsec:sysq-task-types}.

The overlap matrix construction was done using a hierarchical
representation of the basis set, where each part of the hierarchy
contains basis functions located in a particular part of space. At
higher levels in the hierarchy, chunk identifiers are stored referring to basis set
descriptions at lower levels. Using such a hierarchical basis set
description, it is straightforward to implement tasks to compute the
overlap matrix.

The basis function ordering, which affects the sparsity pattern of the
matrix, was determined based on the spatial coordinates of the basis
functions using a recursive divide-space procedure. This is the
default ordering used in the {\sc Ergo} program.

The {\sc Ergo} overlap matrix test calculations were performed on the
Tintin cluster at the UPPMAX computer center, Uppsala University,
using CHT-MPI~1.1 compiled with Open~MPI~1.8.1 and gcc~4.9.1. The AMD Core Math
Library (ACML) version 5.2.0 was used for BLAS operations on
submatrices within the block-sparse leaf matrix implementation.
The Tintin cluster consists of 160 compute nodes. Each node is a
dual AMD Bulldozer compute server with two 8-core Opteron 6220
processors running at 3.0 GHz, with 64 GB of memory. The
nodes are interconnected with a 2:1 oversubscribed QDR InfiniBand
fabric.
CHT-MPI was configured to use 15 threads for executing tasks, leaving
one core on each node to handle communication. The chunk cache size
was set to 8 GB.

The test molecules were water clusters generated from a
molecular dynamics simulation of bulk water at standard temperature
and pressure by including all water molecules within spheres of
varying radii. The Gaussian basis set STO-3G was used, corresponding
to 7 basis functions for each water molecule.
The largest test system contained 1745413 water molecules, giving
12217891 basis functions.
The overlap matrix $S$ was truncated so that the Frobenius norm of the
error matrix was smaller than $10^{-5}$.
The calculations were performed in double precision with leaf matrix
dimension 4096 and blocksize 16.
For the largest test systems this gave a sparsity corresponding to on
average 1070 matrix elements per row in $S$ after truncation, and
about 7000 matrix elements per row in $S^2$.

\begin{figure}
  \begin{center}
    \includegraphics[width=0.48\textwidth]{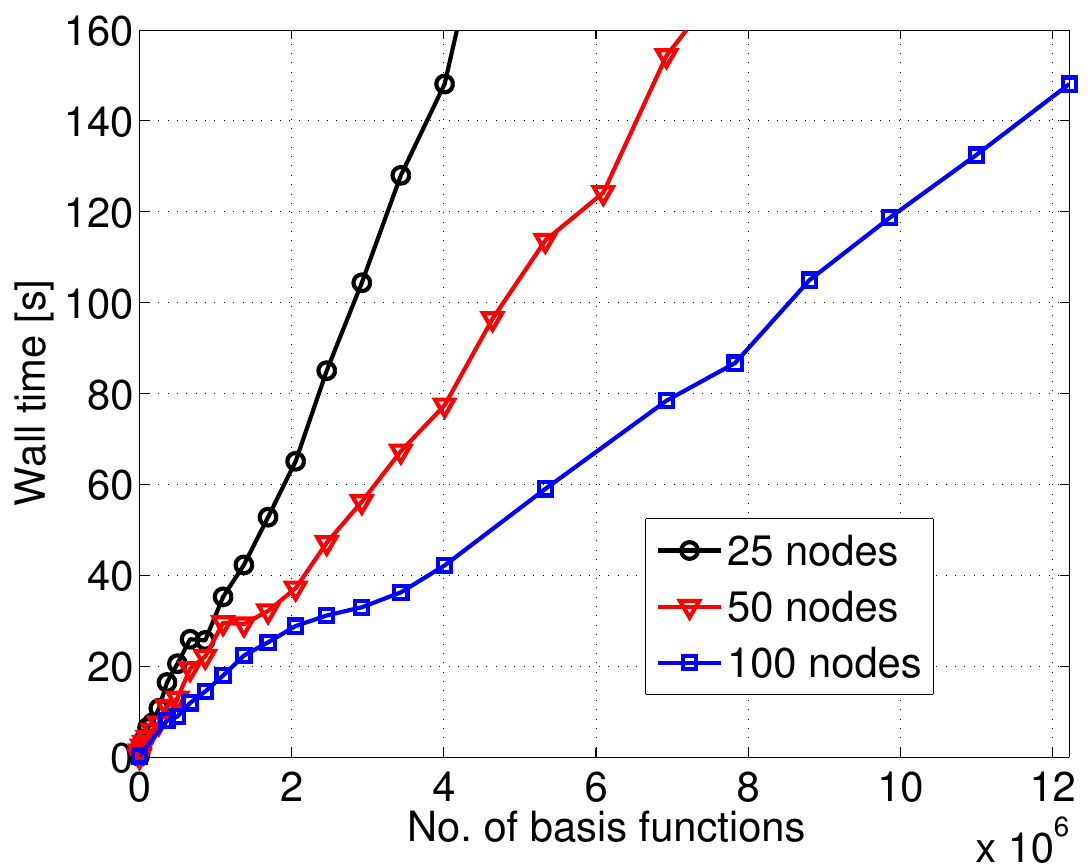}
    \includegraphics[width=0.48\textwidth]{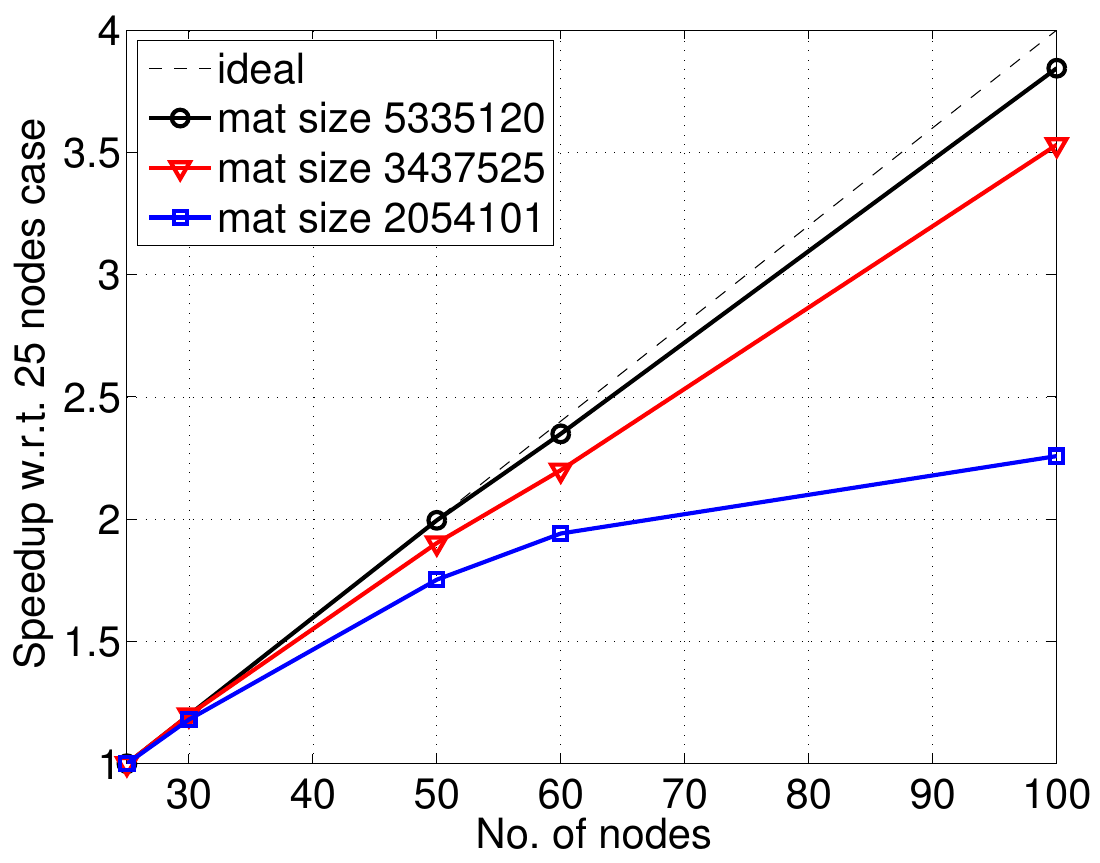}
  \end{center}
  \caption{Timings and scaling for $S^2$ symmetric matrix square computations on
    Tintin.  Left: Timings for $S^2$ computations on
    overlap matrices for water clusters of varying size, using 25, 50,
    and 100 nodes of the Tintin cluster. Nearly linear
    system-size scaling is observed. Right: Scaling with respect to
    number of nodes, for three different matrix sizes. The speedups are
    relative to the 25 nodes case. We get closer to ideal
    speedup when the matrix size is increased.
    \label{fig:smat_tintin_sysq_timings_and_scaling_tintin}}
\end{figure}

\begin{figure}
  \begin{center}
    \includegraphics[width=0.48\textwidth]{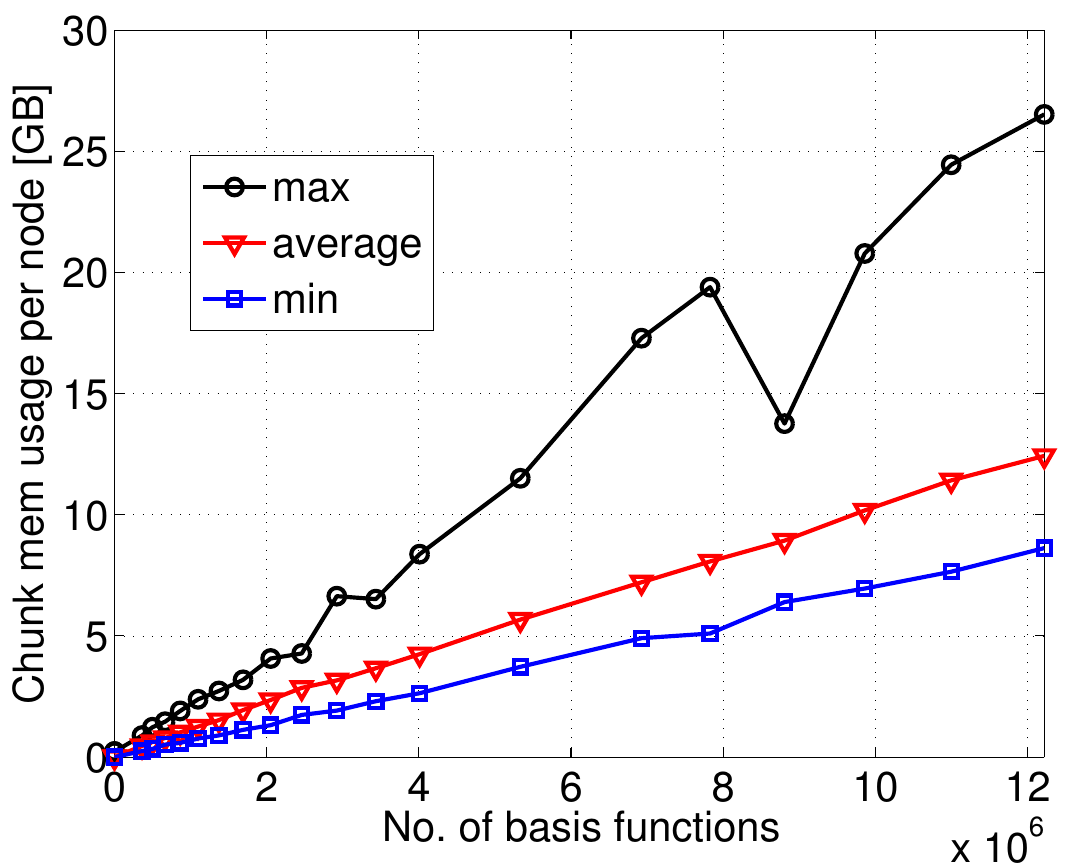}
    \includegraphics[width=0.48\textwidth]{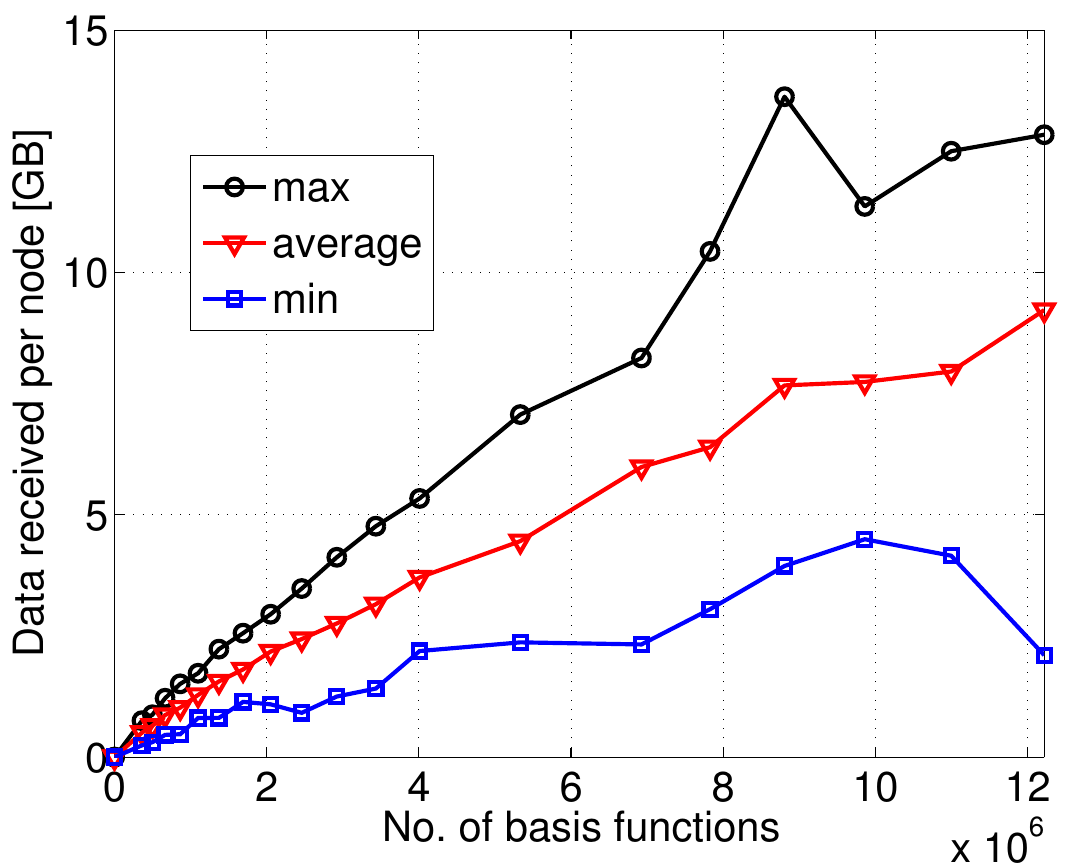}
  \end{center}
  \caption{Memory usage and communication statistics for $S^2$ symmetric matrix square computations
    for water clusters of varying size, using 100 nodes of
    the Tintin cluster. Left: Chunk storage peak memory usage. Only
    the memory for owned chunks is shown here; chunk cache is not
    included.  Right: Amount of data received by each node during the
    symmetric matrix square operation.
    \label{fig:smat_tintin_sysq_memusage_mm}}
\end{figure}

Timings and scaling for different numbers of compute nodes for the
computation of $S^2$ using the symmetric matrix square operation are shown
in Figure~\ref{fig:smat_tintin_sysq_timings_and_scaling_tintin}.
The time scales nearly linearly with the size of the
molecular system, and the parallelization speedup improves for larger
problem sizes.

Figure~\ref{fig:smat_tintin_sysq_memusage_mm} shows memory usage and
communication statistics for the same water cluster $S^2$
calculations. The minimum, maximum, and average values among the 100
compute nodes are shown. Note that since CHT-MPI distributes both work
and data dynamically, both the amount of data stored and the amount of
communication needed will in general be different among the compute nodes.
For the largest water cluster, the $S^2$
matrix contained about $8.6 \times 10^{10}$ matrix elements. Since only the
upper triangle was computed and double precision was used, this
corresponds to 344 GB of storage for $S^2$, or about 3.4 GB per node
for the 100 nodes case. The average chunk memory usage shown in the
left panel of Figure~\ref{fig:smat_tintin_sysq_memusage_mm} is larger,
about 12.4 GB per node, as it includes also temporary matrix chunks
used during the computation.

Compared to the $S^2$ test calculations in
\cite{chunks-and-tasks}, where plain dense matrix storage with
blocksize 500 was used at the lowest level, the results in the present
work
represent significant improvements. The block-sparse leaf matrix type
allows us to exploit sparsity much better, and using the symmetric
matrix square operation reduces the computational effort even further.
For a given water cluster size and number of compute nodes used, the
memory usage for $S^2$ is reduced by about a factor of 16, and the 
time for the $S^2$ computation is reduced by about a factor of 6.
Thanks to the reduced memory usage we are able to test significantly
larger systems.

The results in this section demonstrate that our block-sparse
matrix-matrix multiplication is indeed well suited for applications in
large-scale electronic structure calculations; we get the desired
linear scaling with respect to the size of the molecular system and
reasonable parallel speedup for large enough problems, with dynamic distribution of both work
and data. However, there is room for performance improvements.
Statistics from the calculations indicate that
the worker threads were typically idle more than half of the time, either
waiting for data to be fetched from other nodes or because there was
not enough remaining work to occupy all worker threads. This can be addressed
by improvements within the CHT-MPI implementation, for example by
running tasks closer to their input chunks.
As seen in the left panel of Figure~\ref{fig:smat_tintin_sysq_memusage_mm}, 
a more even distribution of the chunk storage in the the CHT-MPI
implementation would also be desirable, for example using an upper
limit for the chunk storage on each node, and storing chunks elsewhere
when that limit is reached.
Such improvements could be taken advantage of
without changes in the matrix library code, by linking to an
improved CHT-MPI or another Chunks and Tasks library.

The compute nodes on the Tintin cluster where the {\sc Ergo} tests
were run are not equipped with GPUs, so the effect of using GPUs was
not studied here. However, as seen in
Section~\ref{sec:gpu-calculations}, our GPU implementation can provide
additional performance in case GPUs are available. 

Note that the {\sc Ergo} test calculations presented here only involved the overlap
matrix. Full Hartree--Fock or Kohn--Sham density functional theory
calculations require additional parts of the {\sc Ergo} code, notably
the Coulomb and Hartree--Fock exchange matrix construction steps, to
be parallelized using the Chunks and Tasks model. When that is done,
the matrix library described here will be used to combine the
different parts so that full Hartree--Fock and Kohn--Sham density
functional theory calculations can be performed. The most
performance-critical matrix operations are expected to be the
symmetric matrix square operations during density matrix construction.
The density matrix in general contains significantly more nonzero
elements than the overlap matrix~\cite{sparsity-JCC:JCC21723}. Therefore, for a given size of the
molecular system more work for each matrix-matrix multiplication can be expected compared to the $S^2$ tests
here, especially if larger basis sets are also used.

\subsection{Investigation of communication costs}\label{subsec:communication}

As shown in Section~\ref{sec:quadtree-effects}, the quadtree
representation in principle allows data locality in sparse matrices to
be exploited. Here we test this in practice by considering weak
scaling using banded matrices similarly to the calculations presented
in Figure~\ref{fig:weak_scaling_erik} with fixed bandwidth but a
matrix dimension that is increasing proportionally to the number of
worker processes.  The average amount of communication for each worker
process can then be expected to be constant, provided that the used
CHT-MPI library succeeds in distributing the work.

However, if an approach like~\cite{BulucGilbert2012, Borstnik2014} is
used, where a random permutation destroying data locality is employed,
the average amount of communication needed will grow, as noted in
Section~\ref{subsec:CompAndCommCosts}.  In this case, since the matrix
dimension is increased together with the number of processes, by
\eqref{eq:sqrtp_equation}, the number of matrix elements that each
process needs to fetch becomes
\begin{equation} \label{eq:sqrtp_eq2}
  2mk\sqrt{p}
\end{equation}
where $k$ is the constant relating $N$ and $p$ in our weak scaling
tests such that $N = kp$.

Since we are here interested in how the amount of communication
scales for large numbers of processes, we use 8 worker processes per
node and one worker thread per process. 
The test runs used up to 30 nodes of the Tintin cluster, corresponding
to up to 240 worker processes.
We consider here the total amount of data received by each process
from other processes, including both communication between processes
on the same node and between processes on different nodes.
The chunk cache size for each process was set to 2 GB.
The calculations were performed in double precision with leaf matrix
dimension 4096 and blocksize 16.

\begin{figure}
  \begin{center}
    \includegraphics[width=0.48\textwidth]{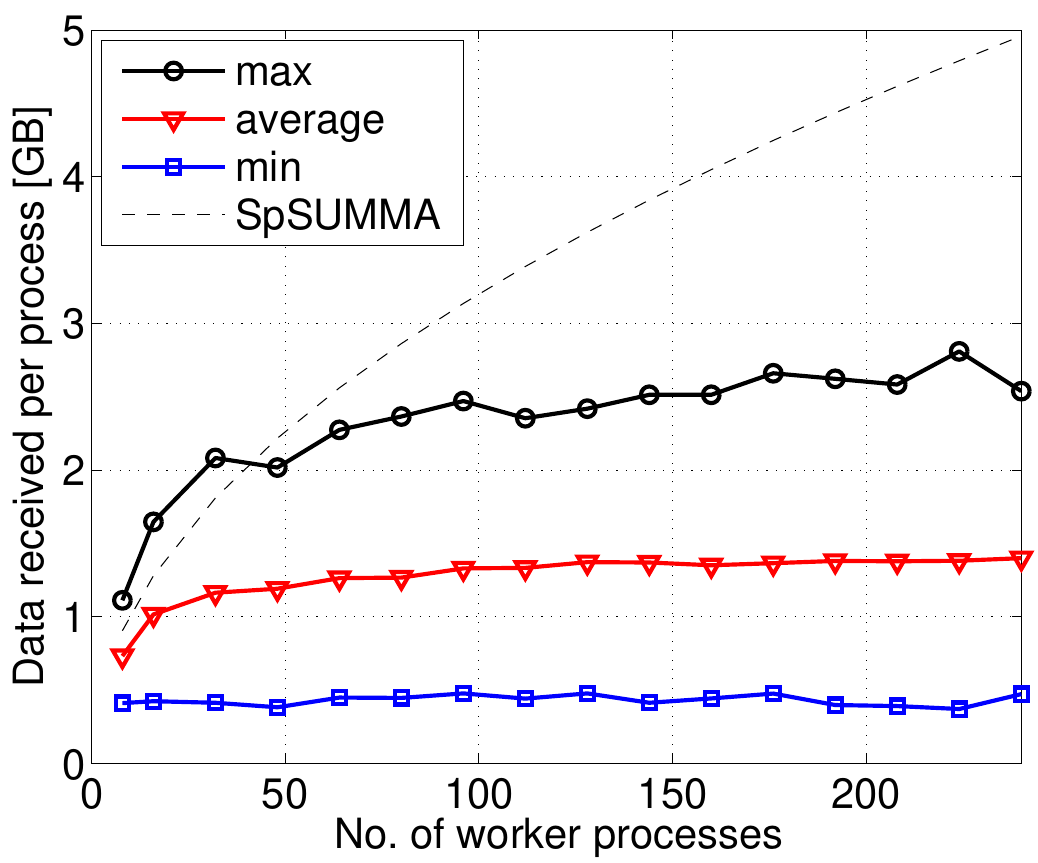}
    \includegraphics[width=0.48\textwidth]{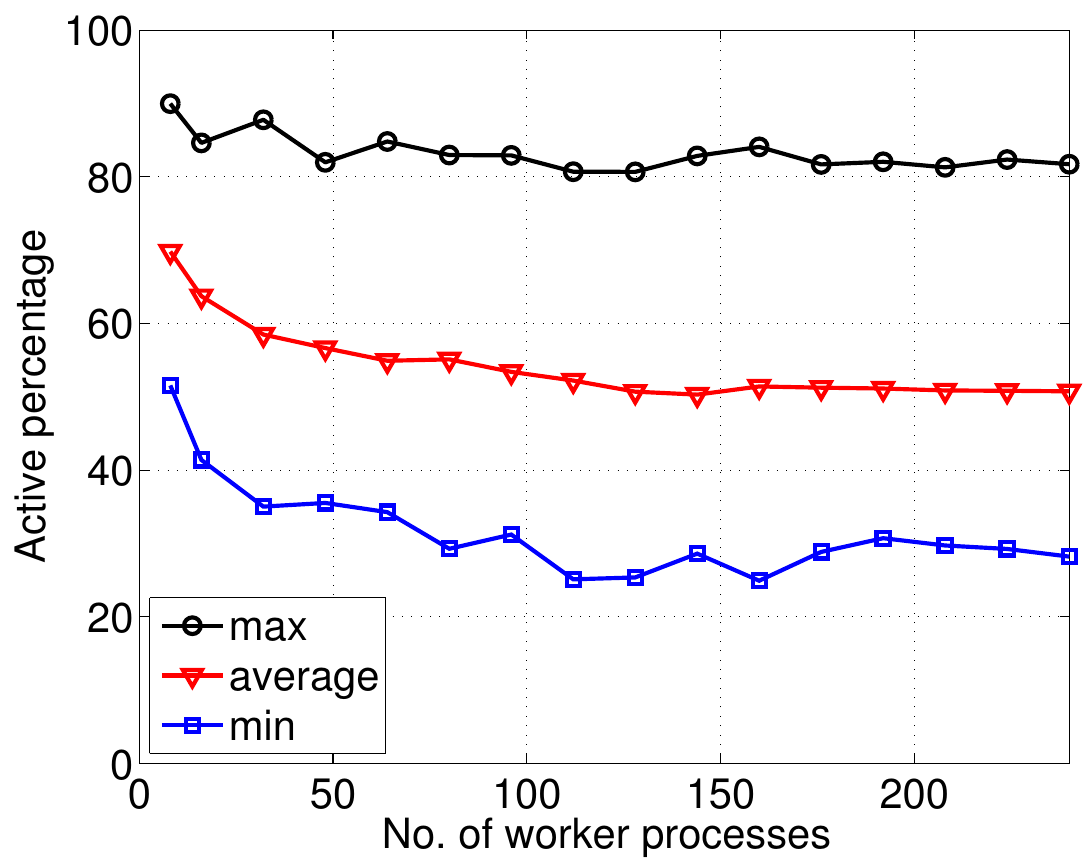}
  \end{center}
  \caption{Weak scaling results for regular matrix-matrix
    multiplication of banded matrices with bandwidth $2 \times 2000+1$
    and matrix dimension $5000p \times 5000p$, where $p$ is the number
    of worker processes. Left: Amount of data received
    by each process during the matrix-matrix multiplication
    operation. The dashed line indicates the amount of communication
    that would have resulted if a random reordering and the Sparse
    SUMMA algorithm had been used, see~(\ref{eq:sqrtp_eq2}).
    Right: active percentage: the fraction of the time that worker
    threads were busy executing tasks.
    \label{fig:tintin_weakscal_regular}}
\end{figure}

Figure~\ref{fig:tintin_weakscal_regular} shows results of our weak
scaling tests using the Chunks and Tasks matrix library.  The minimum,
maximum, and average values among the worker processes are shown. For
each case, the plotted numbers are averages from 6 repeated benchmark
calculations.  For comparison, the amount of communication that would
have been needed if a random reordering and the Sparse SUMMA algorithm
had been used is also shown.  We note that the Chunks and Tasks matrix
library, without a priori knowledge about the sparsity structure, is
able to take advantage of locality and achieve essentially constant
amount of communication per worker process on average, as predicted in
Section~\ref{subsec:CompAndCommCosts}.  To get an indication of the
load balance, the active percentage for the worker threads, defined as
the fraction of the time that worker threads were busy executing
tasks, is also shown in the figure.  When the communication per
process no longer increases the active percentage is also
stabilized. Note the difference between our locality-aware approach
and the Sparse SUMMA algorithm, for which communication would have
dominated completely for large enough test cases.

\begin{figure}
  \begin{center}
    \includegraphics[width=0.48\textwidth]{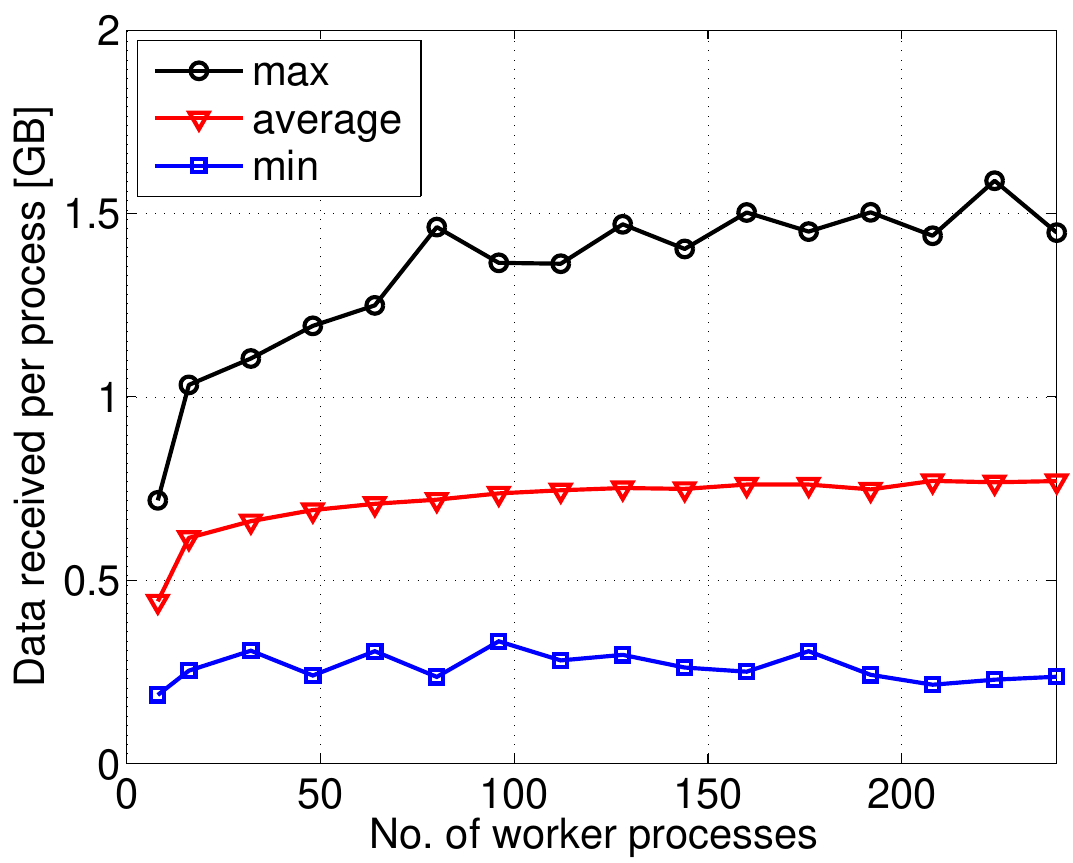}
    \includegraphics[width=0.48\textwidth]{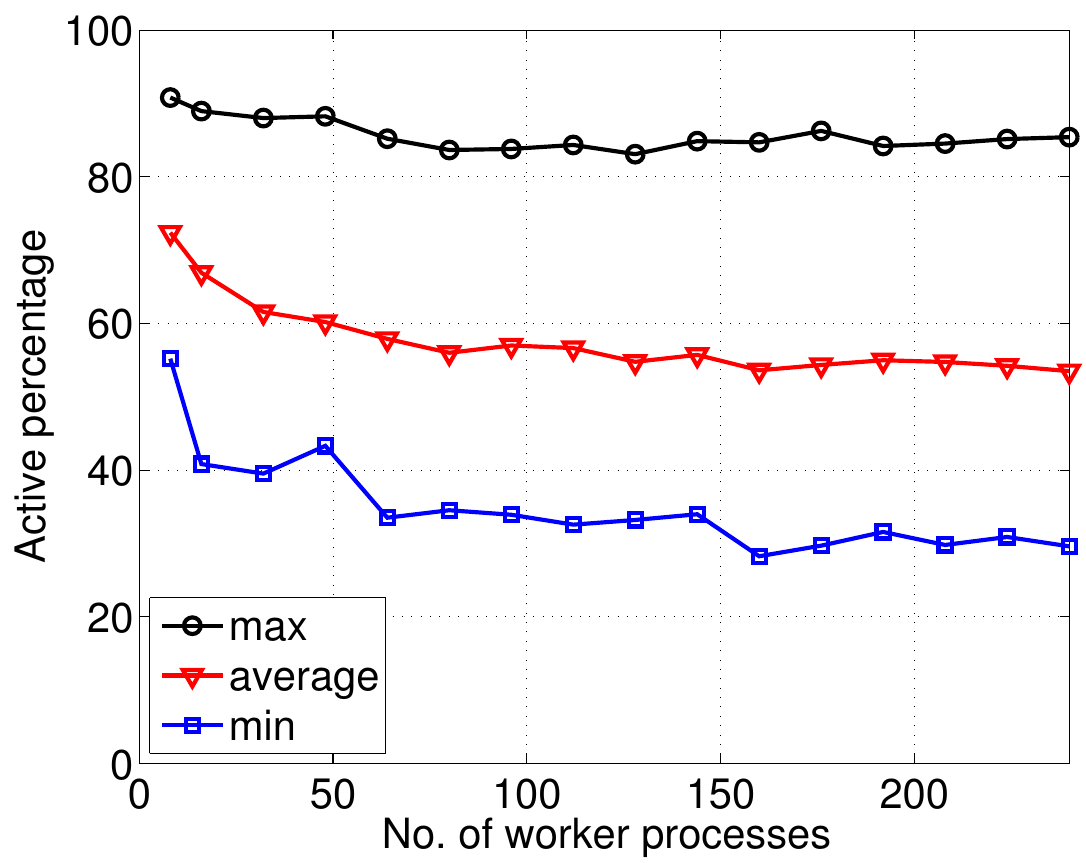}
  \end{center}
  \caption{Weak scaling results for symmetric matrix square
    computations corresponding to the regular matrix-matrix
    multiplication results in
    Figure~\ref{fig:tintin_weakscal_regular}. See that figure caption
    for more information.
    \label{fig:tintin_weakscal_sysq}}
\end{figure}

Figure~\ref{fig:tintin_weakscal_sysq} shows the corresponding results
for the symmetric matrix square operation. For large numbers of
processes, using the symmetric matrix square operation instead of
regular multiplication reduced the average necessary communication per process
from about 1.39 GB to 0.76 GB.

Note that although the weak scaling tests described above were
performed for the simple case of banded matrices, our Chunks and Tasks
matrix library automatically takes advantage of any sparsity and
locality that can be exploited by the quadtree structure.
As another example, exploitation of data locality in the
sparsity pattern improves the scaling of the amount of communication
also for the more complex sparsity patterns occurring in electronic
structure calculations for three-dimensional molecular systems. This can be seen by comparing some of the
overlap matrix tests in Section~\ref{sec:smat-tests}. For example,
going from 2463377 basis functions and 25 nodes to 9861383 basis
functions and 100 nodes corresponds to a factor of 4 in the number of
nodes and a factor of 4.003 in matrix size. The average amount of
data received per node increased from 6.0 GB to 7.7 GB, or a factor of
1.28. This is a significant improvement compared to the factor of 2
that would have resulted from a $\sqrt{p}$ behavior.
It should be noted that such a comparison of different $S^2$
calculations does not correspond exactly to a weak scaling study,
since the spherical shape of the water cluster systems and the use of Frobenius norm truncation lead to an amount of
work that increases slightly more than linearly; in this case the increase
in matrix size by a factor of 4 lead to an increase in the number of
nonzeros in $S^2$ by a factor of 4.33.

\begin{figure}
  \begin{center}
    \includegraphics[width=0.48\textwidth]{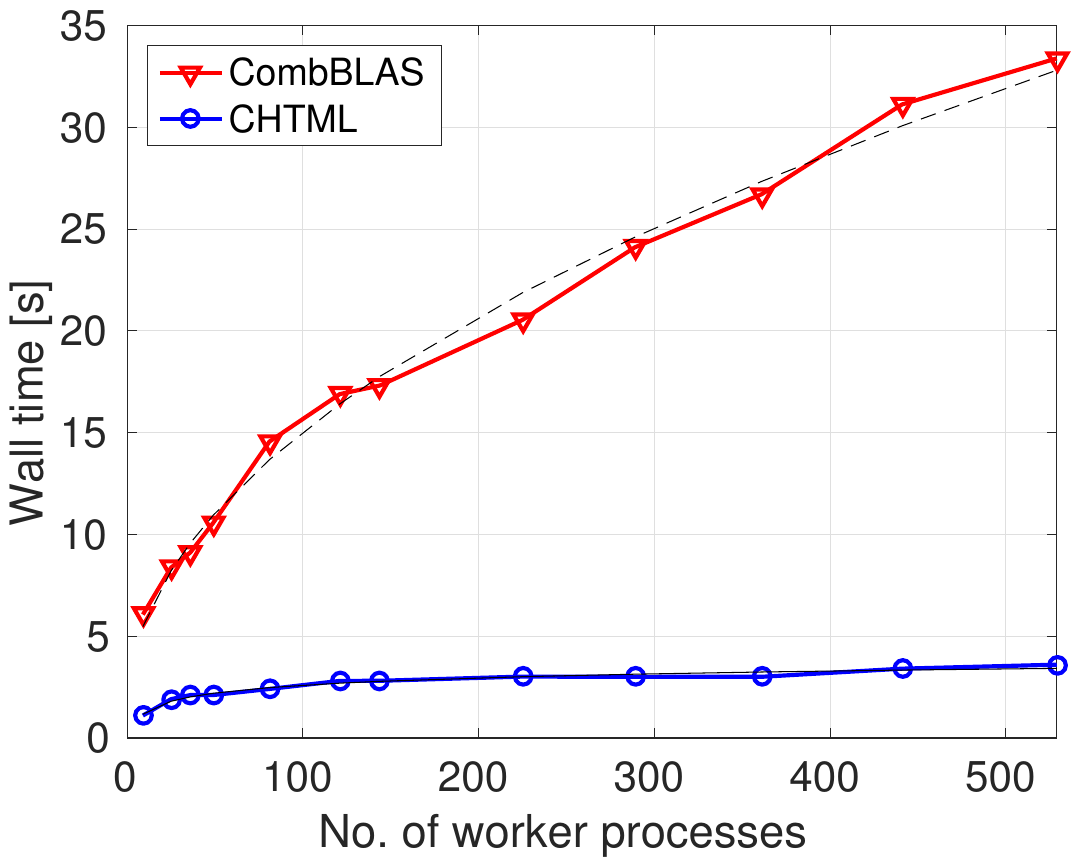}
    \includegraphics[width=0.48\textwidth]{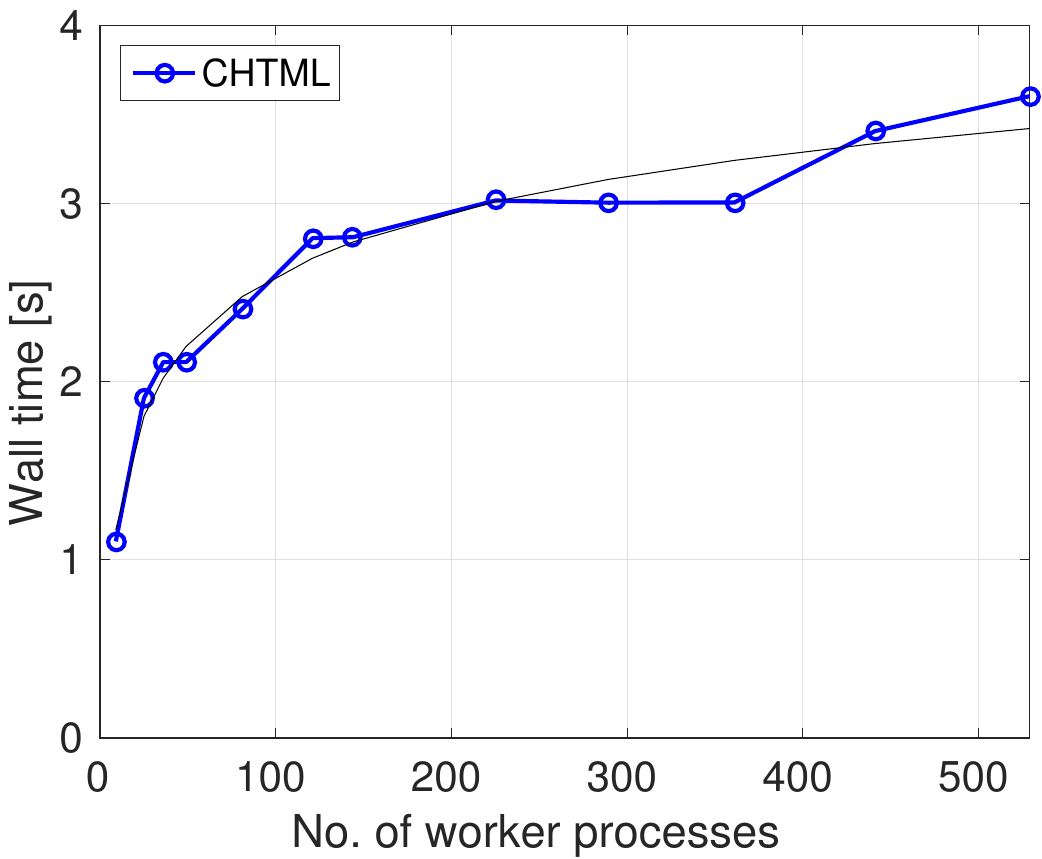}
  \end{center}
  \caption{Timings for weak scaling tests for regular matrix-matrix
    multiplication of banded matrices with bandwidth $2 \times 20+1$
    and matrix dimension $20000p \times 20000p$, where $p$ is the
    number of worker processes. Left: Wall times for Combinatorial
    BLAS (CombBLAS) compared to the Chunks and Tasks matrix library
    (CHTML). Right: Closeup of the CHTML timings.  The dashed and
    solid help lines show $c_0 + c_1 \sqrt{p}$ and $c_0 + c_1 \log(p)
    + c_2 (\log(p))^2$ least squares fits for CombBLAS and CHTML,
    respectively.
    \label{fig:tintin_combblas_comparison}}
\end{figure}

Figure~\ref{fig:tintin_combblas_comparison} shows measured wall times
for weak scaling tests on the Tintin cluster using the Combinatorial
BLAS SpSUMMA implementation~\cite{CombBLAS} and the here presented
matrix library, in order to verify the theoretical results in
Section~\ref{subsec:CompAndCommCosts} regarding execution times.  The
SpSUMMA tests were performed using the freely available and well
organized software package Combinatorial BLAS
version~1.4.0~\cite{CombBLAS}.  Both Combinatorial BLAS and our matrix
library were compiled with Open MPI 1.8.1 and gcc 4.9.1.
The calculations used up to 16 worker processes per node, with the
specific number of processes for each case fulfilling the
Combinatorial BLAS requirement of a square logical processor grid. For
our library, leaf matrix dimension 4096 and blocksize 16 was used and
BLAS operations at leaf level were performed using ACML version~5.2.0.
The test matrices were chosen so that the asymptotic scaling behavior
for each approach can be clearly seen, using banded matrices of size
$N = 20000p$ and bandwidth $2\times 20+1$, in double precision.
This makes the amount of work per process
$\mathcal{O}(d^2N/p)$ relatively small.
For SpSUMMA the $\mathcal{O}(\sqrt{p})$ communication cost dominates
the calculation. SpSUMMA also involves a cost for processing the data
that has been fetched. Owing to the efficient representation of
hypersparse matrices provided by the doubly compressed sparse column
data structure \cite{dcsc2008}, used in Combinatorial BLAS, this cost
is only $\mathcal{O}(\sqrt{p})$ as well. (In contrast, the standard
compressed sparse column representation used in
e.g. Matlab~\cite{matlab_csc1992} and Csparse~\cite{csparse2006} would
lead to an $\mathcal{O}(p)$ scaling for the processing of fetched
data.)  The inherent $\mathcal{O}(\sqrt{p})$ scaling behavior has also been
demonstrated for SpSUMMA as implemented in the distributed
block-compressed sparse row library, see e.g.~Figure~10
in~\cite{Borstnik2014}.
For our quadtree based approach
Figure~\ref{fig:tintin_combblas_comparison} shows a $c_0 + c_1 \log(p)
+ c_2 (\log(p))^2)$ least squares fit to the observed timings. We have
noted that the $c_2$ coefficient is very small and even slightly
negative, meaning that the $\log(p)$ term dominates in practice for
this case. Thus, the observed behavior is slightly better than
predicted by theory, see~\eqref{eq:exec_time_banded}. A possible
explanation is that although the total number of tasks along the
critical path is $\mathcal{O}((\log(p))^2)$, the number of leaf level
tasks along the critical path is $\mathcal{O}(\log(p))$. When a large
blocksize is used, leaf level tasks will be more expensive with
respect to both computation and communication.

In summary, the results of our numerical experiments are consistent
with the theoretical weak scaling results in
Section~\ref{subsec:CompAndCommCosts}: we observe constant average
communication per process and execution times increasing only with the
squared logarithm (or better) for matrices with data locality.  In
such cases, our approach will thus be increasingly favorable compared
to SpSUMMA as the calculation size is scaled up.

\section{Concluding remarks}\label{sec:conclusions}

The matrix library presented in this work is based on a quadtree data
structure, with the important property that it allows for automatic
exploitation of a priori unknown
matrix sparsity structure.  The
hierarchical quadtree representation and associated recursive
algorithms have been implemented using the Chunks and Tasks
programming model.
In this matrix library code, parallelism is exposed to the Chunks and
Tasks library by expressing matrices and operations as hierarchies of
chunk and task objects. All details regarding message passing and
synchronization are left to the Chunks and Tasks library. Thus, the
matrix library code can be written without worrying about how many
nodes there are, where data is to be sent, etc.
Storage and manipulation of matrices at the lowest level of the
hierarchy is handled by a separate leaf matrix library. This means
that the matrix library code is relieved from the details regarding
the best way to store a particular type of submatrix 
or the best way to perform submatrix-submatrix
multiplication on a particular type of hardware.
Such modular design is powerful since it allows each part to be
developed and optimized separately, and one can easily switch between
different implementations of each part. Well designed interfaces
between different modules, in our experience, results in both
increased programming productivity and improved performance.

For matrices appearing in electronic structure calculations, the basis
function ordering plays an important role in determining the sparsity
pattern. For the overlap matrix test calculations in the present
work, the default ordering in the {\sc Ergo} program was used.
A different ordering, using e.g.~space filling
curves~\cite{Challacombe-sparsematrix} or network modularity
optimization~\cite{Girvan11062002-network,rubensson-inverse-factorization},
could lead to increased data locality and result in improved
performance.

In Section~\ref{subsec:CompAndCommCosts} we showed that for matrices
with data locality our quadtree based approach yields superior weak
scaling compared to the Sparse SUMMA
algorithm~\cite{SparseSUMMA2008}. It should be noted that there are
algorithms that in theory could provide better scaling than SpSUMMA,
see for example the so-called 3D algorithms that were theoretically
discussed in~\cite{communication_optimal_2}.  Recently, a practical
implementation of such a 3D algorithm was
demonstrated~\cite{Azad2015}.  In that work, similarly to SpSUMMA, a
random permutation of rows and columns is used to achieve load
balancing.  However, instead of the two-dimensional $ \sqrt{p} \times
\sqrt{p}$ process grid used in SpSUMMA the algorithm makes use of a
three-dimensional $c \times \sqrt{p/c} \times \sqrt{p/c}$ process grid.
Following~\cite{communication_optimal_2}, $c$ should be chosen as
$c=\lceil p/m^2 \rceil$ in order to minimize communication, where $m$
is the average number of nonzeros per row. This makes the 3D algorithm
equivalent to SpSUMMA ($c = 1$) whenever $p \leq m^2$.  This means
that for all benchmark calculations presented in this article, where
$p$ is always smaller than $m^2$, the scaling behavior for the 3D
algorithm would in practice be identical to the scaling behavior of
SpSUMMA. For example, for our tests in Figure~14, where $m = 41$, the
3D algorithm would scale as SpSUMMA up to about 1600 processes.
Although after this point the scaling behavior is expected to be
improved compared to SpSUMMA, the method is still fundamentally
different from our approach since it uses a predetermined data
distribution and is unable to exploit data locality.

The possibility of using quadtrees in data-parallel or partitioned
global address space languages such as High Performance Fortran (HPF)
was discussed by Chatterjee et al.~in
\cite{ChatterjeeMatMul2002_A}. Chatterjee et al.~argued that
representation of dense matrices using quadtrees could be incorporated
in HPF but that recursion and nested dynamic spawning of computations
would be difficult to achieve. In the context of the present work,
representation of sparse matrices with a priori unknown sparsity
patterns would represent a further obstacle.

This work illustrates the usefulness of programming models allowing dynamic distribution of
work and data, which we expect will become increasingly important in the future, as
larger compute systems are used.
Apart from simplifying the implementation of dynamic algorithms, 
such models also make it easier to achieve
fault resilience. They also facilitate the use of heterogeneous
computational resources and allow robustness with respect to varying
performance among compute nodes.

In the test calculations in this work, both generation of input
matrices and verification of output matrices was performed using
Chunks and Tasks programs. Thus, the data distribution of input
matrices was a result of the task executions that generated those
matrices. The placement of chunks was a result of work stealing as
discussed in Section~\ref{subsec:cht-lib-impl}. Using the Chunks and
Tasks matrix library together with other kinds of parallel software
would require conversion of the data structures. Task types for
carrying out such conversion without need for centralized administration
of all data could be provided, although this was not done in the
present work.

Finally, we note that the flexible design of the presented Chunks and
Tasks matrix library makes it easy to modify and extend the
library. For example, inclusion of extra information such as the
Frobenius norm of each submatrix in the quadtree as
in~\cite{BockChallacombeSISC2013} could be straightforwardly
implemented. The library could also be used for elementwise sparse
matrices rather than block-sparse matrices by simply switching to a
different leaf matrix library, employing for example a compressed
sparse row format.

\section*{Acknowledgements}

Support from the G{\"o}ran Gustafsson foundation, the Swedish research
council (grant no. 623-2009-803 and 621-2012-3861), the Lisa and
Carl--Gustav Esseen foundation, and the Swedish national strategic
e-science research program (eSSENCE) is gratefully acknowledged.
Computational resources were provided by the Swedish National
Infrastructure for Computing (SNIC) at the Center for Scientific and
Technical computing at Lund University (LUNARC) and Uppsala
Multidisciplinary Center for Advanced Computational Science (UPPMAX).

\section*{References}

\bibliography{biblio}

\end{document}